\documentclass[12pt,preprint]{aastex}

\shorttitle{The AGN and Star Formation contribution}
\shortauthors{Mel\'endez et al.}

\begin{document}
\title{Constraining  the AGN  Contribution in a Multiwavelength Study  of Seyfert Galaxies}
\author{M. Mel\'endez, S.B. Kraemer}
\affil{Institute for Astrophysics and Computational
Sciences, Department of Physics, The Catholic University of America,
 Washington, DC}
\email{07melendez@cua.edu}
\author{H.R. Schmitt}
\affil{Remote Sensing Division, Naval Research Laboratory, Washington, DC}
\and
\affil{Interferometrics, Inc., Herndon, VA}
\author{D.M. Crenshaw}
\affil{Department of Physics and Astronomy, Georgia State University, Atlanta, GA}

\author{R.P. Deo}
\affil{Department of Physics, Drexel University, Philadelphia, PA}
\author{R.F. Mushotzky}
\affil{NASA/Goddard Space Flight Center, Greenbelt , MD}
\and
\author{F.C. Bruhweiler}
\affil{Institute for Astrophysics and Computational
Sciences, Department of Physics, The Catholic University of America,
 Washington, DC}
\begin{abstract}
We  have studied the relationship between  the high- and low-ionization [O~IV]~$\lambda$25.89~$\micron$, [Ne~III]~$\lambda$15.56~$\micron$  and  [Ne~II]~$\lambda$12.81~$\micron$ emission lines with the aim of constraining the active galactic nuclei (AGN) and star formation contributions for a sample of 103 Seyfert galaxies. We used 
 the [O~IV] and [Ne~II] emission as  tracers for the AGN power and star formation  to  investigate  the ionization state of the emission-line gas. We find  that Seyfert 2 galaxies have, on average, lower [O~IV]/[Ne~II]  ratios than those of Seyfert 1 galaxies. This result suggests two possible scenarios: 1) Seyfert 2 galaxies have intrinsically weaker AGN, or 2) Seyfert 2 galaxies have  relatively higher star formation rates than Seyfert 1  galaxies. We   estimate  the fraction of [Ne~II] directly associated with the AGN and  find that Seyfert 2 galaxies have  a larger contribution from star formation, by a factor of $\sim 1.5$ on average, than what is found in Seyfert 1 galaxies. Using the stellar component  of [Ne~II] as a tracer of the current star formation we found similar  star formation rates in Seyfert 1 
and Seyfert 2 galaxies. We examined the mid- and far-infrared continua  and  find that [Ne~II] is well correlated with the continuum luminosity at 60$\micron$ and that both [Ne~III] and [O~IV] are better correlated with the 25$\micron$ luminosities than with the continuum at longer wavelengths, suggesting that the mid-infrared continuum luminosity is dominated by the AGN, while  the far-infrared  luminosity is dominated by star formation. Overall, these results test the unified model of AGN, and suggest that the differences between Seyfert galaxies cannot be  solely due to  viewing angle dependence.
\end{abstract}

\section{Introduction}

Active Galactic Nuclei (AGN) are thought to harbor massive black holes surrounded by an accretion disk  responsible for the enormous energy rates observed 
in their unresolved nuclei \citep{1984ARA&A..22..471R,2000ApJ...540L..13P,2004ApJ...613..682P}. Historically, Seyfert 1 and Seyfert 2 galaxies 
have been classified by the presence or absence of broad optical emission lines. In this regard, Seyfert 1 galaxies
have  broad permitted  (FWHM$\sim$ 1-5$\times{\rm 10^3~km~s^{-1}}$) and narrow (FWHM$\sim$ 5$\times{\rm 10^2~km~s^{-1}}$) permitted and forbidden lines and  Seyfert 2 galaxies have only narrow permitted and forbidden
 line emission \citep{1974ApJ...192..581K}.   
Using spectropolarimetry, \cite{1985ApJ...297..621A} found broad Balmer lines and [Fe~II]  emission in the polarized spectrum of the Seyfert 2 \objectname{NGC 1068} galaxy, 
characteristic of a Seyfert 1 spectrum. This represents the first observational evidence in favor of a unified model. In this model, Seyfert 1 and Seyfert 2 galaxies are intrinsically the same with their differences attributed to viewing angle. In Seyfert 2 galaxies, our line of sight to the broad line region (BLR) and the central engine  is obstructed by an optically thick  dusty torus-like structure, while in Seyfert 1 galaxies, our line of sight is not obstructed by the torus, allowing 
a direct view of the central regions of the active galaxy.

Although it has been observationally confirmed, the unified model for Seyfert galaxies does not address  the role of stellar activity. The
 fact that active galaxies can  also host massive star-forming regions
 \citep[e.g,][]{1990MNRAS.242..271T,1997ApJ...485..552M,1998ApJ...493..650M,2004MNRAS.355..273C,2006MNRAS.366..480G,2007ApJ...671.1388D} suggests a connection between the AGN and star formation in the proximity of the super-massive black hole, typically on scales of a few hundred parsecs. These starbursts may have a significant impact on the fueling of the central black hole \citep[e.g.,][]{1999MNRAS.303..173S,2007ApJ...671.1388D}. Many authors have suggested 
that violent star formation, in the circumnuclear region,  plays a fundamental role in the energetics of Seyfert 2 galaxies \citep[e.g.,][]{1985MNRAS.213..841T,1998ApJ...493..650M,1995MNRAS.272..423C,2001ApJ...546..845G,2001ApJ...558...81C}. This extra source of energy, a young  stellar population in the vicinity of the nucleus,  could 
 complement  the non-stellar component associated with  the AGN. \cite{1997ApJ...485..552M}
 suggested that  asymmetric morphologies and bars, especially in Seyfert 2 galaxies, are an important factor in  star formation and Seyfert classification. These  
asymmetric morphologies can induce radial gas inflow and fueling of the nuclear region, thus  obscuring and  feeding  the active nucleus. \cite{2001ApJ...546..845G} discuss
the possibility of two kinds of Seyfert 2 galaxies based on their stellar population properties: those with young and intermediate age stars and those with 
the optical continuum dominated by old stars.

 In principle, the richness of the infrared spectrum provides a unique opportunity
 to test the unified model of AGN, since  the mid- and far-infrared spectra appear to be different in Seyfert 1 and Seyfert 2 galaxies \citep[e.g.,][]{1988ApJ...328L..35S,1992ApJ...401...99P,1993ARA&A..31..473A,2000A&A...357..839C,2007ApJ...656..148A}. However, there is  the technical 
difficulty of  isolating the AGN  from  contamination by the host galaxy emission and, more importantly, star formation features \citep[e.g.,][]{2004A&A...418..465L,2005ApJ...633..706W}. In this work we will focus on deconvolving the different contributions (e.g., AGN+star formation) in the
 [Ne~II]~$\lambda$12.81~$\micron$ emission line and the mid- and far-infrared continua. In order to estimate the  component associated with the AGN  we will use the high-ionization potential ($\sim {\rm 54.9~eV}$) [O~IV]~$\lambda$25.89~$\micron$ emission line. We   found \citep[][hereafter M08]{2008arXiv0804.1147M} a tight correlation in 
Seyfert 1 galaxies between the [O~IV] and the X-ray 14-195~keV continuum luminosities from  {\it Swift}/BAT observations \citep{2005ApJ...633L..77M}. A weaker 
 correlation was  found in Seyfert 2 galaxies, which was due to   partial absorption in the 14-195~keV band. Overall, we proposed  
[O~IV] as a truly isotropic property of AGNs given its high ionization potential and that is basically unaffected by reddening, meaning that 
the [O~IV] strength directly measures the AGN power. In this work we will  isolate the    stellar component of the [Ne~II] emission to trace the 
instantaneous star formation rates in our sample of Seyfert galaxies.


\section{The Infrared Sample\label{sample}}

Our sample of Seyfert galaxies includes  compilations from \cite{2007ApJ...671..124D}, \cite{2007arXiv0710.4448T}, \cite{2002A&A...393..821S},  \cite{2005ApJ...633..706W}  and the X-ray selected sample from M08. This sample  has been  expanded to include 
[O~IV]~$\lambda$25.89~$\micron$, [Ne~II]~$\lambda$12.81~$\micron$  and [Ne~III]~$\lambda$15.56~$\micron$  fluxes   from  our analysis of  unpublished 
archival spectra observed with the Infrared Spectrograph (IRS) \cite[see][]{2004ApJS..154...18H}
 on board  {\it Spitzer} in the ${\rm 1^{st}}$ Long-Low 
(LL1, $\lambda$ = 19.5 - 38.0 $\mu$m, 10.7$'$$'$ $\times$ 168$'$$'$, $R\sim60-127$), Short-High (SH, $\lambda$ = 9.9 - 19.6 $\mu$m, 4.7$'$$'$ $\times$ 11.3$'$$'$,$R\sim 600$) and Long-High (LH, $\lambda$ = 18.7 - 37.2 $\mu$m, 11.1$'$$'$ $\times$ 22.3$'$$'$, $R\sim600$) IRS order   in  Staring mode. The sample includes 64 Seyfert~2 and 39 Seyfert~1
 galaxies which are listed in Table~\ref{ir}. The infrared luminosities 
are presented without reddening corrections. For comparison, we searched the literature for the Seyfert 1 and Seyfert 2 
galaxies with measured  mid- and far-infrared 
continuum fluxes at 25$\micron$, 60$\micron$ and 100$\micron$ from the 
{\it Infrared Astronomical Satellite} ({\it IRAS}) \citep{1984ApJ...278L...1N,1989AJ.....98..766S,1990BAAS...22Q1325M,2003AJ....126.1607S}. Note that it was not possible to find {\it IRAS} fluxes for all galaxies in the sample.  For the analysis of 
the mid-infrared emission lines observed with IRS/{\it Spitzer} we followed  the procedure described in M08.

\begin{figure}
\epsscale{.80}
\plotone{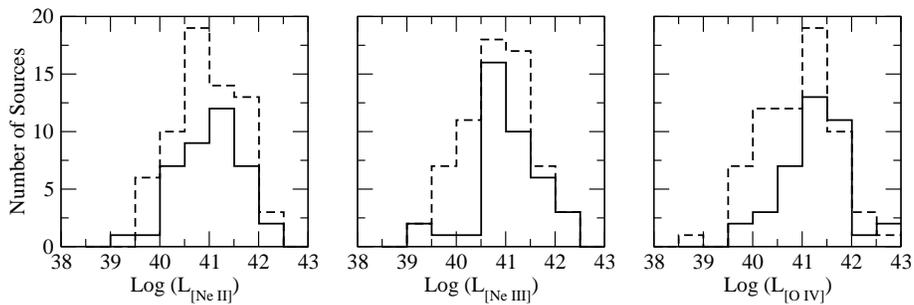}
    \caption{Comparison of the  [Ne~II], [Ne~III] and [O~IV] luminosities  in Seyfert 1 ({\it solid line}) and Seyfert 2 ({\it dashed line}) galaxies. This sample includes 39 Seyfert~1 and 65 Seyfert~2 galaxies.
The K-S test for these emission line luminosities show that two samples drawn from the same population would differ this much $\sim 79.0\%$, $\sim 22.8\%$ and 
 $\sim 10.0\%$  of the time for the [Ne~II], [Ne~III] and [O~IV] luminosity distributions, respectively. \label{ir_fig1}}
\end{figure}

First, we need to confirm that our sample, which was compiled from various sources, was not biased in terms of luminosity, e.g., toward high luminosity Seyfert 1 and/or low-luminosity Seyfert 2 galaxies. This test was made by comparing the different luminosities  for
both the mid-infrared emission lines and continuum luminosities for the two groups of galaxies. The results from these 
comparisons are shown in Figure~\ref{ir_fig1} with the results from the  Kolmogorov-Smirnov (K-S) test presented in Table~\ref{ks}.
This table also includes 
information about the numbers of Seyfert 1 and Seyfert 2 galaxies, mean values and standard deviations of the mean for the measured quantities.

The histograms of [Ne~II], [Ne~III] and [O~IV]  luminosities are presented in Figure~\ref{ir_fig1}. From the [Ne~II] histogram  it can be seen that Seyfert 1 and Seyfert 2 galaxies have  similar distributions of values.  The K-S test for this emission line luminosity returns a $\sim 79.0\%$ probability of the null hypothesis (i.e.,
 that there is no difference between Seyfert 1 and Seyfert 2 galaxies), or in other words,  two samples drawn from the same parent population would differ this much 
$\sim 79.0\%$ of the time\footnote{A probability  value of less than 5\% represents a high level of significance that two samples drawn from the same population are different. A strong level of significance is obtained for values smaller than 1\% \citep[e.g.,][]{1992nrfa.book.....P,2003drea.book.....B}}. From the distribution
 of [Ne~III] luminosities one  can see  the relative  absence of low luminosity Seyfert 1 as compared with Seyfert 2 galaxies. For $L_{{\rm [Ne~III]}} < 40.5$ there are 
 20 Seyfert 2 galaxies (comprising $\sim 30\%$ of the Seyfert 2 sample) and only four Seyfert 1 galaxies ($\sim 10\%$ of the Seyfert 1 sample). This may  suggest the presence  of intrinsically weaker AGN in the Seyfert 2 galaxies. However,  the K-S result returns $\sim 22.8\%$ probability of the null hypothesis, indicating that the 
apparent differences between the two groups are not statistically significant. From the [O~IV] histogram one  can seen a lack of low luminosity Seyfert 1 ($L_{{\rm [O~IV]}} < 40.5$) as compared with Seyfert 2 galaxies, similar to that found in the distribution of [Ne~III] luminosities, with 20 Seyfert 2 galaxies comprising $\sim 30\%$ of the Seyfert 2 sample and only five Seyfert 1 galaxies ($\sim 13\%$ of the Seyfert 1 sample). However, 
 the K-S result returns a $\sim 10.0\%$ probability of the null hypothesis. Overall,  we found that the mid-infrared luminosity distributions for Seyfert 1 and Seyfert 2 galaxies are statistically similar, even with the absence of low luminosity Seyfert 1 galaxies in the [Ne~III] and [O~IV] distributions. 

We also studied the infrared emission continuum properties of the sample using the  25$\micron$, 60$\micron$, 100$\micron$ and far-infrared (FIR)   luminosities. It should be noted that the large beam infrared spectral energy distributions from {\it IRAS} (with a field of view of  0.75$'$$\times$4.6$'$ at 25$\micron$, 1.5$'$$\times$4.7$'$ at 60$\micron$ and 
3.0$'$$\times$5.0$'$ at 100$\micron$) includes the AGN continuum and  the host galaxy emission \citep{1995ApJ...453..616S,2004A&A...418..465L}. The far-infrared luminosity (FIR) is characterized by the emission at 60~$\mu {\rm m}$ and 100~$\mu {\rm m}$ \citep[e.g.,][]{1992ARA&A..30..575C,1996ARA&A..34..749S}. In this regard, the 60~$\mu {\rm m}$ emission represents a ``warm" component 
associated with dust around young star-forming regions. On the other hand,  cooler ``cirrus" emission at 100~$\mu {\rm m}$ \citep{1984ApJ...278L..19L} 
is associated with  a more extended dust heated by the interstellar radiation field \citep{1998ARA&A..36..189K}.

\begin{figure}
\epsscale{.80}
\plotone{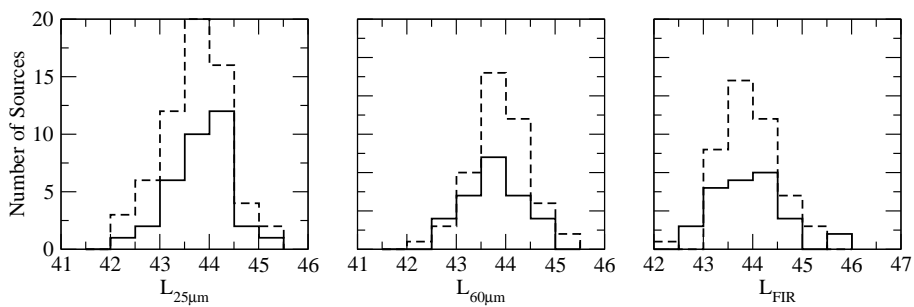}
    \caption{Comparison of the  25$\mu$m, 60$\mu$m and FIR  luminosities  in Seyfert 1 ({\it solid line})  and Seyfert 2 ({\it dashed line}) galaxies. This sample includes 36 Seyfert~1 and 63 Seyfert~2 galaxies.
The K-S test for these continuum luminosities returns $\sim 27.3\%$, $\sim 70.4\%$ and $\sim 51.3\%$ probability of the null hypothesis for the 25$\mu$m, 60$\mu$m and FIR  continuum luminosities distributions, respectively.\label{ir_fig15}}
\end{figure}

The histograms of  $25~\mu{\rm m}$ ($L_{25\mu{\rm m}}$), $60~\mu{\rm m}$ ($L_{60\mu{\rm m}}$) and far-infrared, $FIR~\mu{\rm m}$ ($L_{FIR}$), continuum luminosities  are presented  in Figure~\ref{ir_fig15}. Overall,  it can be seen that Seyfert 1 and Seyfert 2 galaxies have a similar distribution of values. For the $25~\mu{\rm m}$ 
 luminosities distributions the K-S test  returns a $\sim 27.3\%$ probability of the null hypothesis. For the
 $60~\mu{\rm m}$ luminosity ($L_{60\mu{\rm m}}$) the K-S test returns $\sim 70.4\%$ probability of the null hypothesis. This result is in agreement with 
previous studies that have assumed that  the $60~\mu{\rm m}$ continuum emission is an isotropic quantity  \citep[e.g.,][]{2001ApJ...555..663S}. However,  this assumption 
 must be adopted with caution because the  torus emission may be anisotropic at $60~\mu{\rm m}$ \citep{1992ApJ...401...99P}. Figure~\ref{ir_fig15} shows that Seyfert 1 and Seyfert 2 galaxies also have   similar distributions in the FIR. The K-S test returns  a $\sim 51.3\%$ probability of the null hypothesis for the FIR distribution.  

\begin{figure}
\epsscale{.50}
\plotone{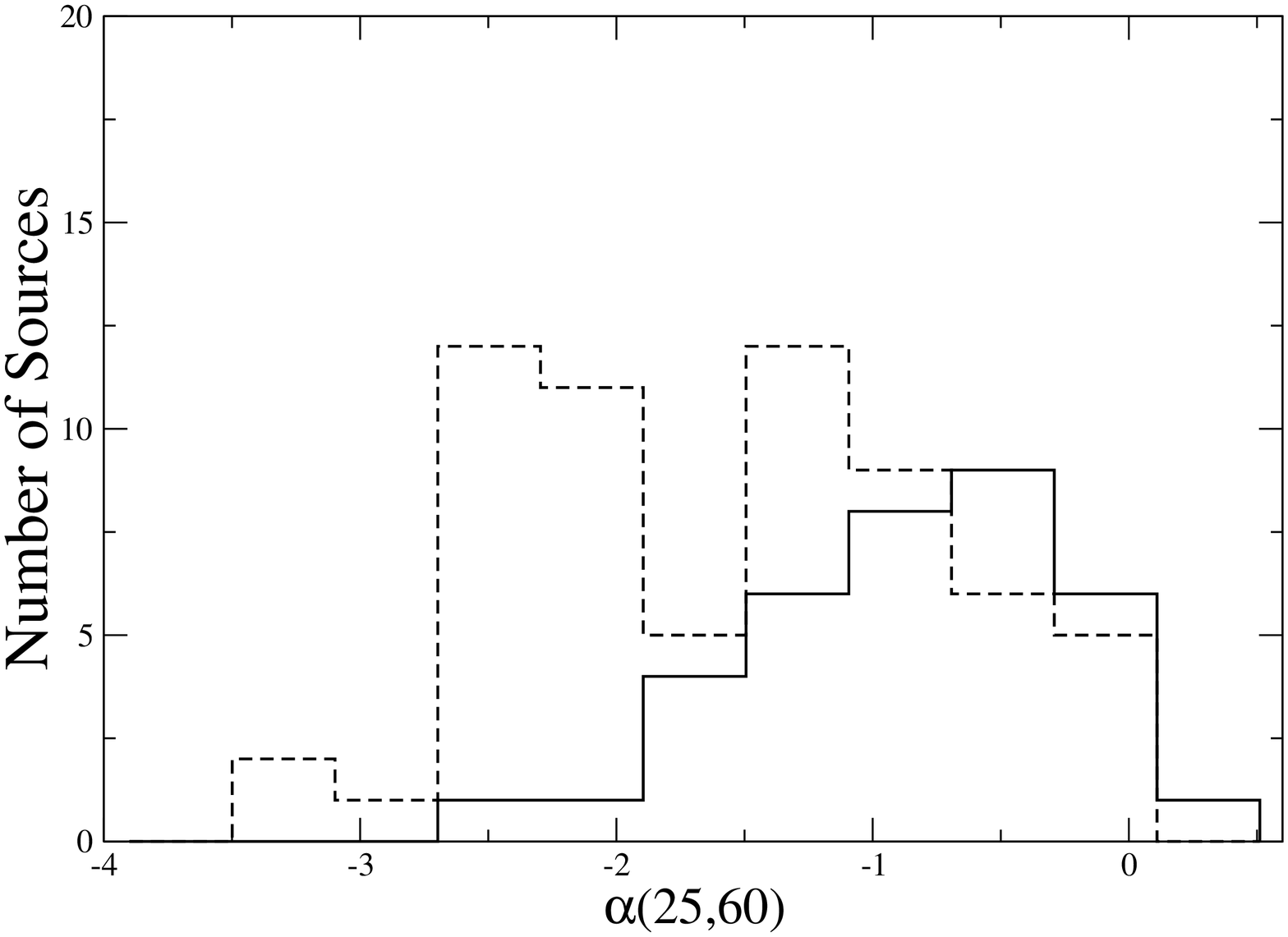}
    \caption{Comparison between the mid- to far-infrared index, $\alpha_{25-60}$, for Seyfert 1 ({\it solid line}) and Seyfert 2 ({\it dashed line}) galaxies. This sample includes 35 Seyfert~1 and 60 Seyfert~2 galaxies.
The K-S test for the spectral index returns $0.1\%$ probability of the null hypothesis.\label{ir_index}}
\end{figure}

In spite of the fact that Seyfert 1 and Seyfert 2 galaxies have similar distributions of infrared luminosities, using the observed mid- and far-infrared continuum fluxes, we found a clear difference in their spectral index\footnote{The continuum is assumed to be a power law, $F_{\nu}\propto \nu^{\alpha}$, where $\alpha$ is the spectral index} $\alpha_{25-60}$ in Seyfert 1 and Seyfert 2 galaxies, as  shown in  Figure~\ref{ir_index}. The K-S test returns $\sim 0.1\%$ probability of the null hypothesis. The spectral shape between  25$\micron$ and $60\micron$ has been used to separate Ultraluminous Infrared Galaxies (ULIRGs) that are ``warm"  and  possibly dominated by the AGN and those that are ``cold"  and likely 
to be dominated by star formation \citep{1988ApJ...328L..35S,2007ApJ...656..148A}. From our sample, we found  that Seyfert 2 galaxies 
 possess   relatively cooler dust, with an average $\alpha_{25-60}=-1.5\pm0.1$,  than Seyfert 1 galaxies, 
$\alpha_{25-60}=-0.8\pm0.1$, in agreement with 
previous findings by \cite{2003ApJ...583..159H}. This result suggests that the infrared spectra of Seyfert 1 galaxies 
are  dominated by   hot dust heated by the AGN \citep[e.g.,][]{1997Natur.385..700H,2001ApJ...546..845G}.

\section{Emission Line Diagnostics}

\begin{figure}
\epsscale{.80}
\plotone{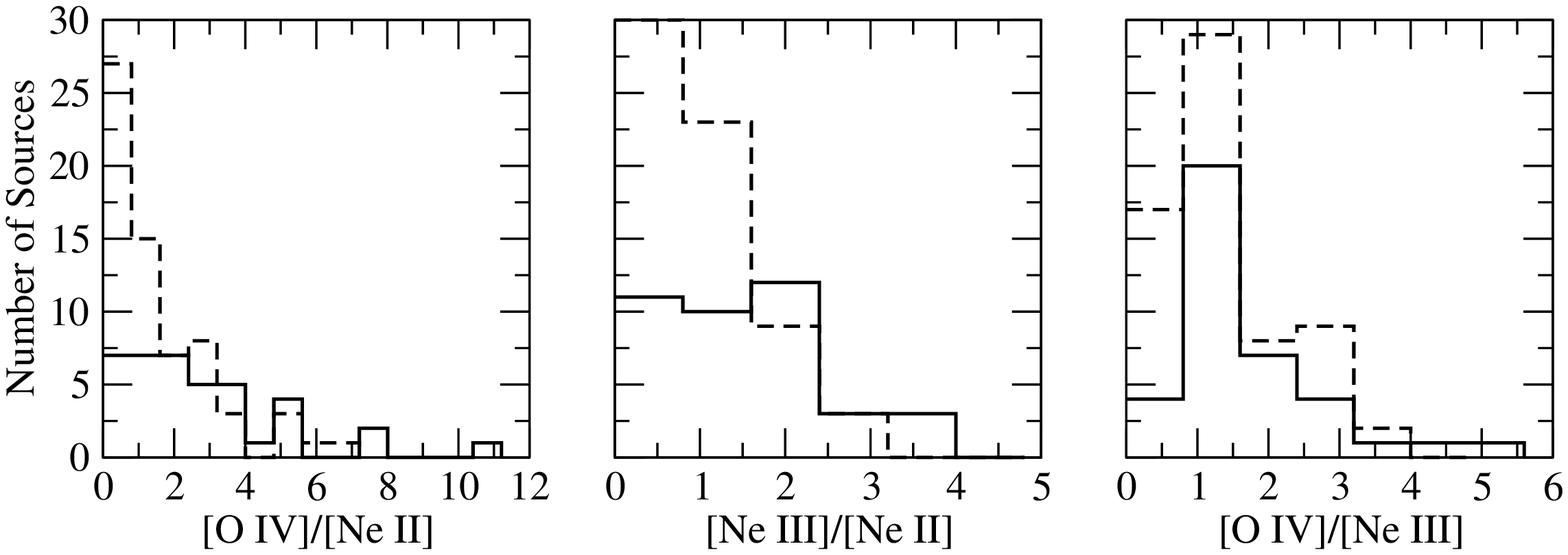}
    \caption{Comparison of the [O~IV]/[Ne~II], [Ne~III]/[Ne~II] and [O~IV]/[Ne~III] distributions in Seyfert 1 and Seyfert 2 galaxies. For the 
[O~IV]/[Ne~II], [Ne~III]/[Ne~II] and [O~IV]/[Ne~III] ratios the K-S test returns  $\sim 0.9\%$, $\sim 1.1\%$ and $\sim 37.7\%$  probability of the null hypothesis, respectively. \label{ir_fig4}}
\end{figure}

The ratios of high- and low-ionization mid-infrared emission lines have been widely used to separate the relative contribution of the  AGN and star formation
\citep[e.g.,][]{1998ApJ...498..579G,2002A&A...393..821S,2006ApJ...646..161D}. We  performed a statistical analysis for the [O~IV]/[Ne~II], [Ne~III]/[Ne~II] and
 [O~IV]/[Ne~III] ratios for  our sample of 103 Seyfert galaxies, with  the results 
presented in Table~\ref{ks}. Figure~\ref{ir_fig4} shows the histograms of  [O~IV]/[Ne~II],
 [O~IV]/[Ne~III] and [Ne~III]/[Ne~II] ratios. From the [O~IV]/[Ne~II]  ratios, it can be seen that   Seyfert 
2 galaxies are displaced toward smaller values than those found for Seyfert 1's, in agreement with 
previous findings by \cite{2007ApJ...671..124D}. Accordingly, in the  sample,
 Seyfert 2 galaxies have, on average, smaller [O~IV]/[Ne~II]  ratios than those observed in Seyfert 1 galaxies. The K-S test returns $\sim 0.9\%$ probability of the null hypothesis, indicating, that  the two Seyfert groups  are statistically different. As
 for  [Ne~III]/[Ne~II], Seyfert 2 galaxies are again displaced toward values smaller than those found for Seyfert 1 galaxies, with the majority of Seyfert 2 galaxies 
($\sim 60 \%$ of the Seyfert 2 population) having  ${\rm [Ne~III]/[Ne~II]} < 1.0 $. The K-S test returns $\sim 1.1\%$ probability of the null hypothesis. From 
the histogram of [O~IV]/[Ne~III],  it can be seen that both 
groups of galaxies have similar distributions. The K-S test returns  a $\sim 37.7\%$ probability of the null hypothesis. This result suggests that  [Ne~III] is also an isotropic quantity and could be used to  estimate AGN power \citep[e.g.,][]{2007ApJ...671..124D,2007ApJ...655L..73G,2007arXiv0710.4448T}.

\begin{figure}
\epsscale{.80}
\plotone{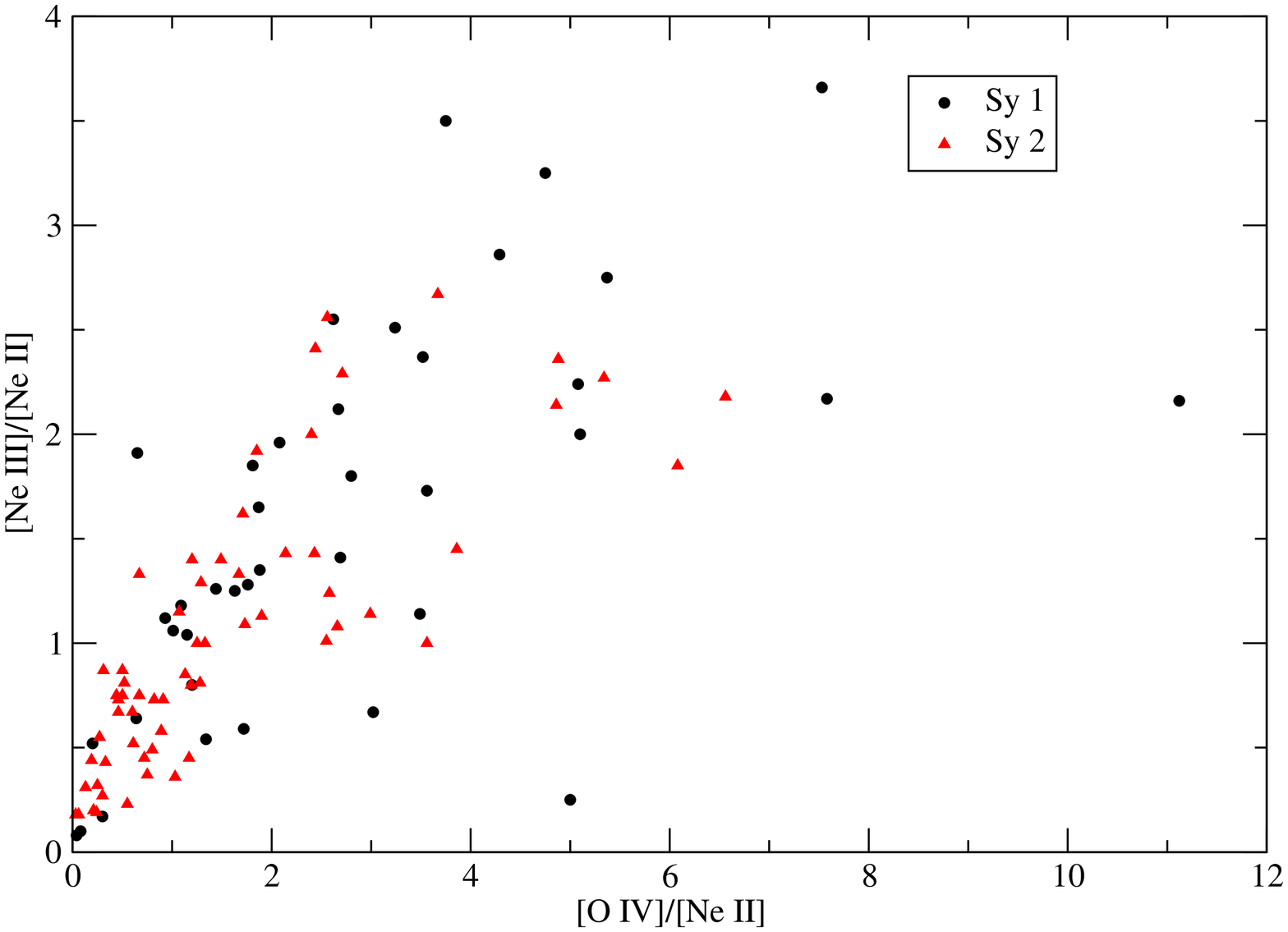}
    \caption{[Ne~III]/[Ne~II] vs.  [O~IV]/[Ne~II] ratios in Seyfert 1 (circles) and Seyfert 2 (triangle) galaxies\label{ratio1}. Of particular interest in this 
comparison is that  below ${\rm [Ne~III]/[Ne~II]} < 1.0 $ and ${\rm [O~IV]/[Ne~II]} < 2.0 $ there are  only 9 Seyfert 1 galaxies but 
 the majority of the Seyfert 2 galaxies ($\sim 60 \%$ of the Seyfert 2 population)}
\end{figure}

In Figure~\ref{ratio1} we compare the [O~IV]/[Ne~II] and  [Ne~III]/[Ne~II] ratios in Seyfert 1 and Seyfert 2 galaxies, where  the  Spearman rank test
 returned  a strong correlation ($r_s=0.810~P_r=1.0 \times 10^{-15}$). This strong 
correlation supports the utility of the [Ne~III]/[Ne~II] ratio as a
  diagnostic of the relative strength of the AGN as found by \cite{2007arXiv0710.4448T}. From this correlation we found that  Seyfert 2 galaxies show
 lower [O~IV]/[Ne~II] and [Ne~III]/[Ne~II] ratios than Seyfert 1 galaxies. As suggested by \cite{2007ApJ...671..124D}, one 
needs to consider  two possible scenarios. 1) Seyfert 1 galaxies have  more highly ionized  narrow line regions (NLR) than Seyfert 2 galaxies resulting in an apparently weaker AGN in  Seyfert 2 galaxies;  2) Seyfert 2 galaxies have a relatively higher star formation rates than Seyfert 1  galaxies normalized 
to the AGN luminosity. In the former scenario, one needs to consider that there could be a source of obscuration on a much larger scale that is affecting the [O~IV] but not the [Ne~II] emission; however, this possibility contradicts the findings of [O~IV] as a true isotropic quantity for the AGN (see M08). Alternatively, 
 this hidden inner NLR scenario is ruled out by \cite{2007ApJ...671..124D} where they found  the amount of extinction in the mid-infrared  for Seyfert 1.8/1.9s
 to be negligible.

We  found that Seyfert 1 and Seyfert 2 galaxies are statistically different  in their relative contribution of the  AGN and star formation, as given by the 
analysis of their [O~IV]/[Ne~II] and [Ne~III]/[Ne~II] ratios. Nevertheless, one needs to check if the  emission-line ratios  are not biased because of the relative absence of low luminosity Seyfert 1 galaxies in our sample. In this regard, the  poor correlation found  ($r_s=0.489 ~P_r=9.7\times 10^{-7}$) between [O~IV]/[Ne~II]  and [O~IV]
 luminosities, and the fact that  the K-S test results for the [O~IV] and [Ne~III] luminosity distribution indicated that both groups have statistically similar 
distributions (see discussion in Section~\ref{sample}), indicates that our emission-line ratios are not biased
 because of the relative absence of low luminosity Seyfert 1 galaxies in our sample.  The [O~IV]/[Ne~II] emission line ratio was also compared with the redshift ($z$) in order to check if our previous results are biased towards small ratios
at higher values of $z$, given the fact that [O~IV] is likely to be produced in a compact region, whereas [Ne~II] could  be produced in a more extended region. 
The Spearman rank $r_s$ test did not show any correlation ($r_s=-0.075,~P_r=0.68$) with $z$. At the median redshift for the sample, $z= 0.02$, 1$'$$'$ represents $\sim$400~pc for $H_o=71~{\rm km}$~${\rm s^{-1}}$~${\rm Mpc^{-1}}$. Hence, for example, the SH slit will sample about $\sim$2~kpc in the dispersion direction at $z$ of 0.02.
  
\section{Deconvolving the  Stellar Contribution to the [Ne~II] emission}
\begin{figure}
\epsscale{.90}
\plotone{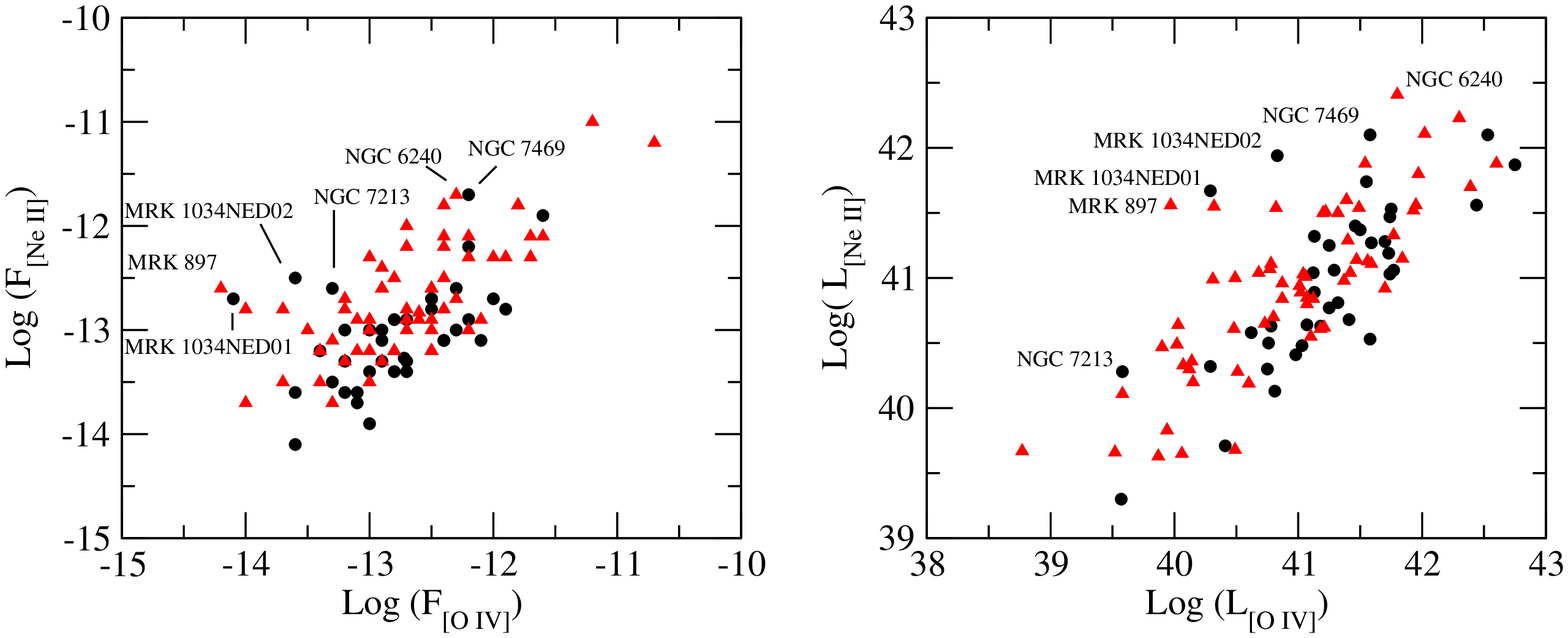}
    \caption{Correlation between [Ne~II] versus [O~IV] fluxes and luminosities. The identified  sources are known to have 
 strong star formation activity. Symbols are identical to those in Figure~\ref{ratio1}. \label{ir_all}}
\end{figure}

In Figure~\ref{ir_all} we compare [Ne~II] and [O~IV] fluxes and luminosities. There appear to be a relative deficiency of Seyfert 1 galaxies in the upper left region of the plot. In the plot we identified   some of 
the outliers in the sample.  All these sources show  strong star formation activity, and have also been  classified as starburst galaxies or to harbor massive 
 H~II regions. These galaxies
 show strong polycyclic aromatic hydrocarbon (PAH) features at 6.2~$\mu$m and 11.5~$\mu$m \citep{2007ApJ...671..124D,2007arXiv0710.4448T}. PAHs are 
a class of large organic  molecules that have been observationally associated with  star formation \citep[e.g.,][]{2000A&A...357..839C,2004A&A...419..501F,2005ApJ...633..871C,2006ApJ...649...79S}. The strong 
[Ne~II]  in these objects  indicates that [Ne~II] is a quantitative tracer of star formation. Results from the statistical analysis are
 presented in Table~\ref{corre1}. One should note that, due to redshift effects, luminosity-luminosity  plots will almost always 
show some correlation. Thus, we are primarily interested in the tightness of the correlations or the slopes (e.g., of one class versus another). 

\begin{figure}
\epsscale{.90}
\plotone{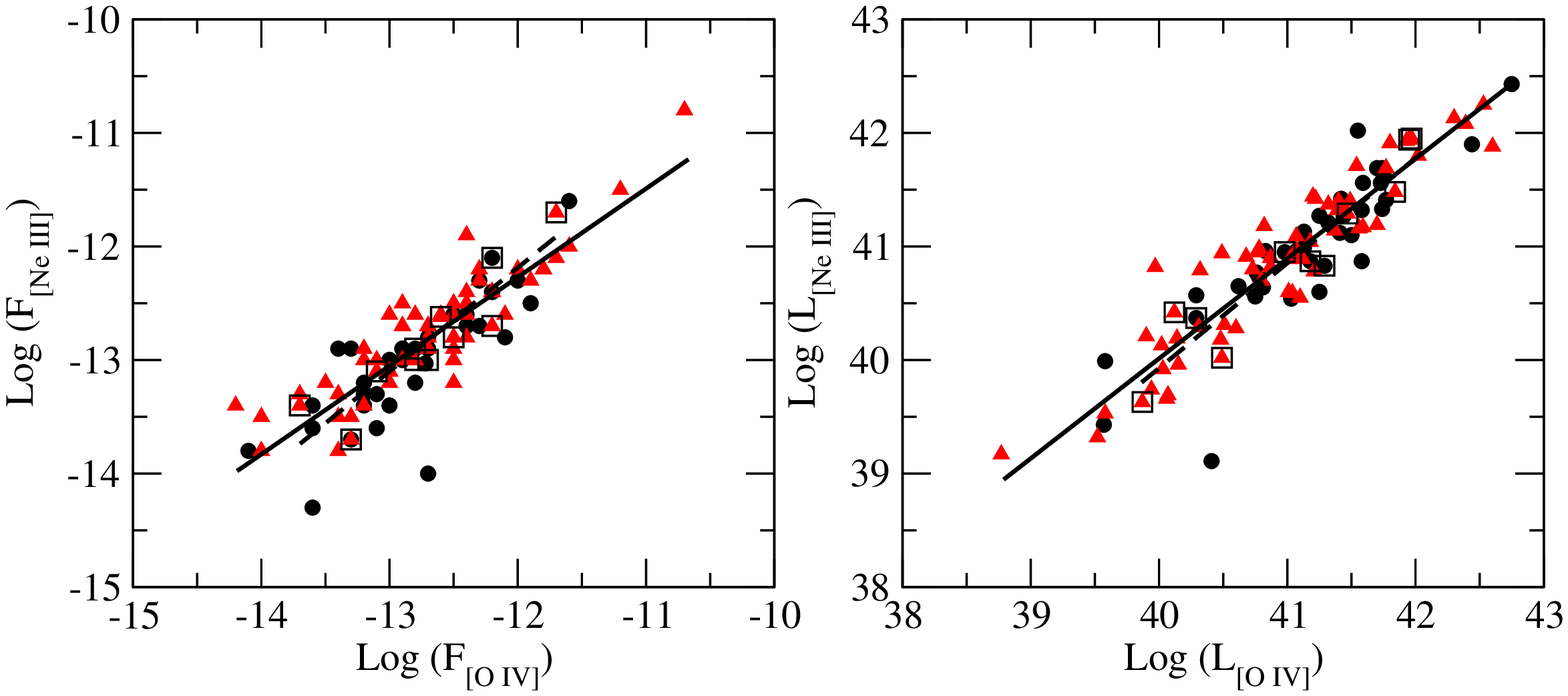}
    \caption{Correlation between [Ne~III] and [O~IV] fluxes, and luminosities. The squares represent  AGN with no detectable PAH features 
at 6.2~$\mu$m and 11.5~$\mu$m. The 
{\it solid line} represents the linear fit obtained for the full sample and the {\it dashed line} represents the linear fit for the pure AGN sources. Symbols are identical to those in Figure~\ref{ratio1}.\label{ir_fig8}}
\end{figure}

Figure~\ref{ir_fig8} shows the correlation between the [O~IV] and [Ne~III] in fluxes (left) and luminosities (right), where 
 a tight and strong correlation is seen (also, see Table~\ref{corre1}). This result corroborates the effectiveness of [Ne~III] as a tracer of the AGN power
  as discussed   by \cite{2007ApJ...655L..73G}, who found a strong correlation between   [Ne~V]$\lambda$14.3~$\micron$ emission-line, which originates 
from an ion with an ionization potential of $\sim 97~{\rm eV}$ and thus is due almost entirely to AGN photoionization \citep{2008ApJ...678..686A},  and [Ne~III].
  Although, there is evidence of [Ne~III] emission from stars \citep[e.g.,][]{2007ApJ...658..314H}, the tight 
correlation between the [O~IV] and [Ne~III] suggest that this contribution is minimal in our AGN sample. In this regard, Figure~\ref{ir_fig8}
 also shows 11  ``pure" AGN sources (see Table~\ref{ir}), i.e., sources that shows  no detectable PAH features at 6.2~$\mu$m and 11.5~$\mu$m in their spectra. 
 One  can see the similarity between  the full sample and the pure AGN linear fits. The linear regression values for the pure AGN sources are 
given in Table~\ref{corre}.

 Assuming that the observed  [Ne~II] emission  is composed of both  an AGN and a stellar component, we estimated the star formation contribution (SC) 
 in the [Ne~II] emission by subtracting  the predicted [Ne~II] that is  produced  by the AGN. In this regard, the predicted [\ion{Ne}{2}] emission associated with the AGN
 was obtained  from the linear regression for the pure AGN sources from  the observed  [Ne~II]~-~[O~IV] correlation in luminosities.

\begin{figure}
\epsscale{.90}
\plotone{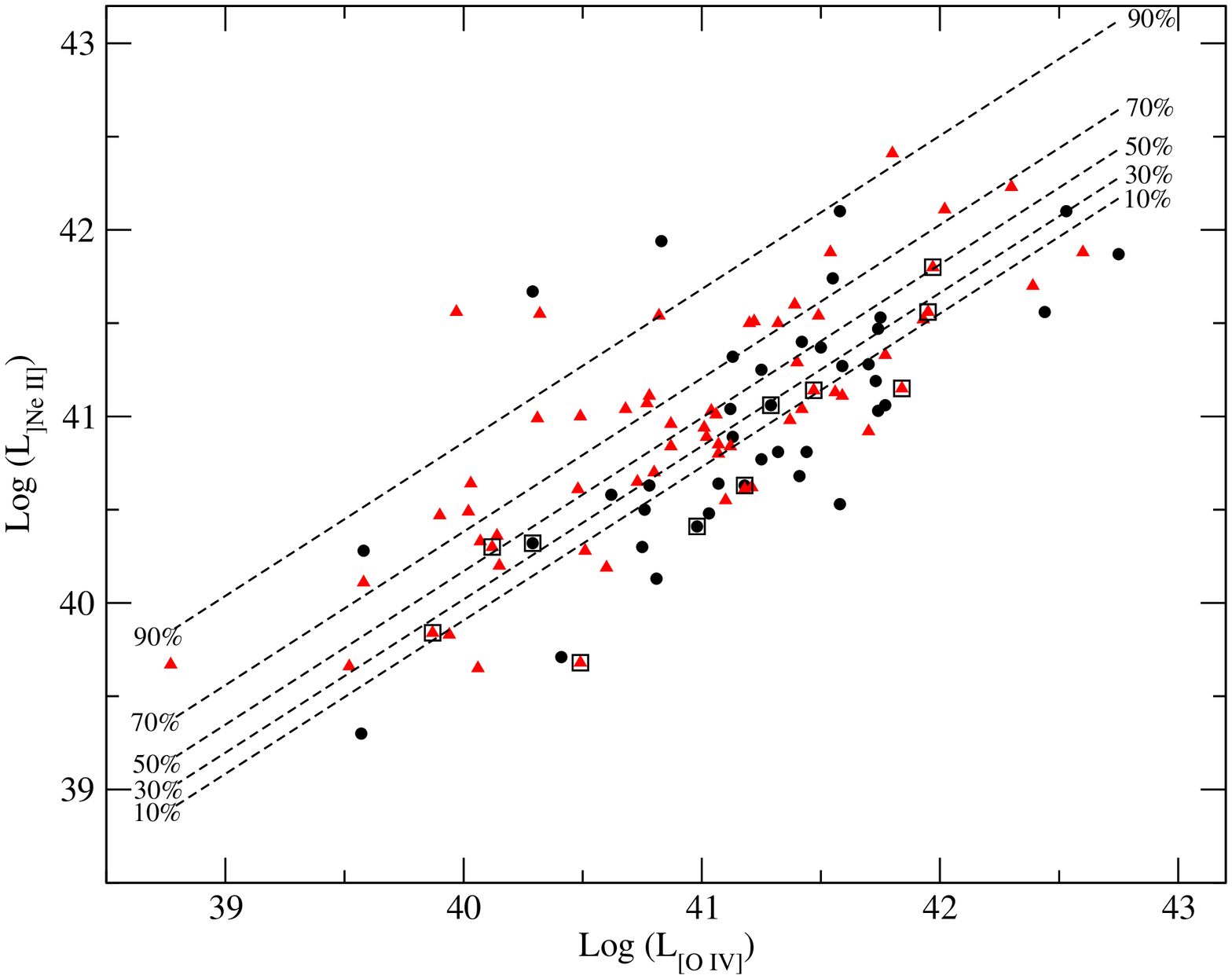}
    \caption{Correlation between [Ne~II] and [O~IV] luminosities. The {\it dashed lines} represent the percentage of stellar component  in the
[Ne~II] emission line, based on the predicted AGN contribution. The squares represent  AGN with no detectable PAH features 
at 6.2~$\mu$m and 11.5~$\mu$m. Symbols are identical to those in Figure~\ref{ratio1}.\label{sc_o4_lum}}
\end{figure}

Figure~\ref{sc_o4_lum} shows the correlation between [Ne~II] and [O~IV]  luminosities. The different dashed lines represent the percentage of 
star formation contribution in the observed [Ne~II] emission lines. Table~\ref{ir} shows the percentage of star formation contribution, SC (\%),  in the [Ne~II] observed luminosity.  For example, NGC~7469 and NGC~3079 are  known to  harbor regions of   enhanced star formation \citep{1996ApJ...468..191P,1998ApJ...498..579G,2005ApJ...633..706W}. In these galaxies, the AGN contributes   $\sim 12\%$ for NGC~7469 and $\sim 18\%$ for NGC~3079
 to the observed [Ne~II] flux. These results are in agreement with the strong PAH 6.2~$\micron$ ($F_{6.2\micron}~ > 5.3 \times 10^{-19}{\rm W~cm^{-2}}$) 
observed in these objects by \cite{2007ApJ...671..124D}.
 \cite{1998ApJ...498..579G} generated an empirical  diagram (mixing model) to separate  
 the AGN from galaxies powered by enhanced star formation, by using the ratio of high- and low-ionization mid-infrared lines and PAH features. We compared their 
results with values obtained  in the present work for our sample. For example, \cite{1998ApJ...498..579G} find that Mrk~273 has
 an AGN contribution of $\sim40\%$, compared with   $\sim34\%$ from our empirical diagnostic.  
 The Seyfert galaxies \objectname{NGC 4151}, \objectname{NGC 1068}, \objectname{NGC 5506} and \objectname{NGC 3783} are AGN dominated, i.e., more than $75\%$ of 
the [Ne~II]'s contribution from the AGN, as suggested by previous mixing models \citep{1998ApJ...498..579G,2002A&A...393..821S}. These results are in excellent agreement with the star formation and AGN contribution  values that we 
 obtained from these objects (see Table~\ref{ir}). The Seyfert 2 galaxy, \objectname{NGC 6240}, is dominated by star formation \citep{1996A&A...315L.137L,1998ApJ...498..579G} in agreement with our value of a $\sim 90\%$ stellar contribution to the [Ne~II] emission line. We estimated an error of $\sim 20\%$ in the  AGN-predicted [Ne~II] luminosities, based on the star formation contribution  obtained for \objectname{Mrk 3}, which is one of the objects with no detectable PAH at 6.2$\mu$m \citep{2007ApJ...671..124D}.

Table~\ref{ks} shows the results from the  statistical analysis for the star formation contribution  in the [Ne~II] emission. We  found that, averaged over populations,  Seyfert 2 galaxies have a stronger stellar contribution in their [\ion{Ne}{2}] observed emission line, $\sim 43\pm4$\%,  than that found in Seyfert 1 galaxies, $\sim 28\pm5$\%. The K-S test returns $\sim 4.4\%$ probability of the null hypothesis, meaning that the two groups of Seyfert galaxies are statistically different in their relative stellar 
contribution to their [Ne~II] emission, despite the fact that there are Seyfert 1 galaxies in our sample with strong starbursts (e.g., NGC~7469). Nevertheless,  these results are 
in agreement with previous findings where Seyfert 2 galaxies typically show stronger starburst signatures  in their infrared spectra than Seyfert 1 
galaxies \citep[e.g.,][]{2006AJ....132..401B}.

\begin{figure}
\epsscale{.90}
\plotone{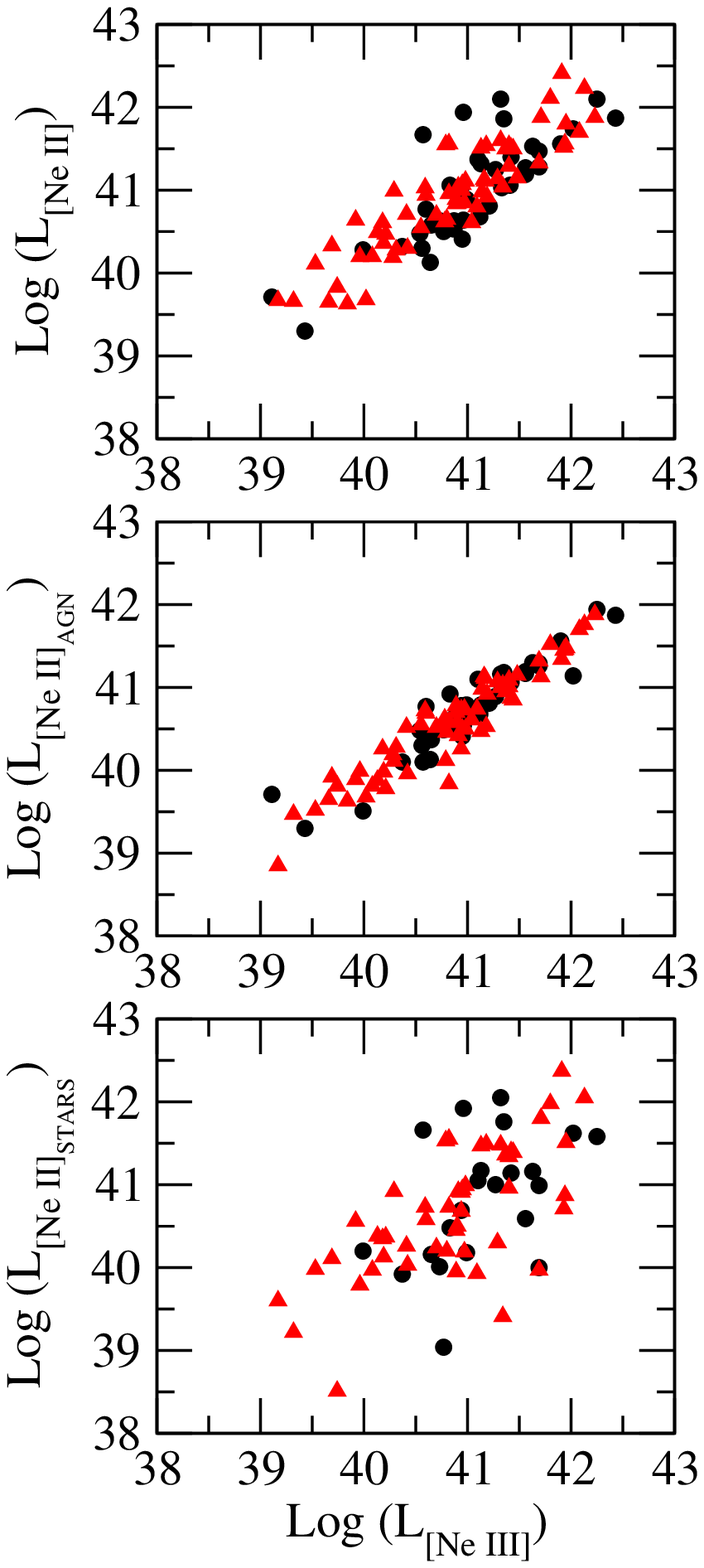}
    \caption{Correlation between [Ne~II] and [Ne~III] after the deconvolution of  the AGN  and stellar  component in the observed [Ne~II]  luminosities. In the upper, middle and lower panel
 we present  the observed (AGN+stellar component), AGN and the stellar component for [Ne~II] versus  [Ne~III]  emission line luminosities. Symbols are identical to those in Figure~\ref{ratio1}.\label{ir_fig111}}
\end{figure}

In Figure~\ref{ir_fig111} we plot the  result from the deconvolution method performed to the observed [Ne~II] emission. In the upper, middle and lower panel
 we present  the observed (AGN+stellar component), the AGN-only, and the stellar component for [Ne~II] versus  [Ne~III]  emission line luminosities.  In order to avoid 
a misleadingly linear correlation, we plotted the observed, stellar and AGN components of [Ne~II] against [Ne~III]. A tight correlation between
   [Ne~III] and the pure AGN [Ne~II] validates the method used to untangle the different contributions  in the
[Ne~II] emission line. In the lower panels of these plots, one note the lack of correlation between the stellar component of [Ne~II] and the strength of the 
AGN, as characterized by the [Ne~III] emission.

\section{Star Formation Rate }

Using a sample of non-AGN star-forming galaxies, \cite{2007ApJ...658..314H}  investigated the utility of the mid-infrared emission lines of [Ne~II] and 
[Ne~III]  as a star formation rate (SFR) indicator, given the fact that the Lyman continuum radiation, which can ionize  ${\rm Ne^+}$ and ${\rm Ne^{2+}}$,  is mainly produced by young stars. 
In order to calculate the SFR we have used the stellar component of [Ne~II], as deconvolved from the previous analysis, and   assumed  $f_+=0.75$ and $f_{+2}=0$ for the fraction of Neon in the form of ${\rm Ne^0}$ and ${\rm Ne^{+}}$, respectively \citep{2007ApJ...658..314H}.  Since we could not extract
 the star formation contribution from the [Ne~III] emission lines, given the tight correlation with [O~IV], we assumed
 that all  the [Ne~III] emission is coming from the AGN. Therefore, the SFR derived only from the [Ne~II] represents a lower limit for the region within the extraction 
aperture for a fixed value of the fractional abundances for  ${\rm Ne^0}$ and ${\rm Ne^{+}}$.  The SFR ($M_\sun{\rm ~yr^{-1}}$) values are presented in Table~\ref{ir}. One must note that the predicted 
star formation rate may depend on the aperture size, accordingly  the last column of  Table~\ref{ir} shows the size (in kpc) that the slit samples in the dispersion direction.

We performed a K-S test analysis for the derived star formation rate for Seyfert 1 and Seyfert 2 galaxies, from this analysis we found  $\sim 18.2\%$ probability of the null hypothesis, suggesting that Seyfert 1 and Seyfert 2 galaxies have statistically similar star formation rates. In this regard
 we found the star formation rate in  Seyfert 2 galaxies to have an average  of  
$ 8 \pm 2 M_\sun{\rm ~yr^{-1}}$ and  $7\pm 2 M_\sun{\rm ~yr^{-1}}$ for Seyfert 1 galaxies. Caution must be taken on the interpretation of the star formation rates derived from the [Ne~II] luminosity. This derived SFR is a probe of the 
young massive stellar population and is independent of the previous star formation history. As an example, \cite{2007ApJ...671.1388D} analysed the star 
formation history in the AGN dominated Seyfert 2 galaxy NGC~1068. They estimated a $SFR\left (M_\sun{\rm yr^{-1}~kpc^{-2}}\right )= 90-170$ with 
a starburst age of 200-300~Myr. Our results are significantly lower, $SFR\left (M_\sun{\rm yr^{-1}~kpc^{-2}}\right )= 0.45$,  suggesting minimal  current star formation,
 in agreement with the extensive analysis discussed by \cite{2007ApJ...671.1388D}. In general, our results are systematically lower than 
that found by \cite{2007ApJ...671.1388D}, given the fact that in the nuclear regions of their sample of Seyfert galaxies there appears to have been recent, 10-300~Myr,
 starbursts  that must have already ceased.  In other words, their diagnostic sample contained an older stellar population than that mapped by the [Ne~II] which  only 
traces  young ($<{\rm 10~Myr}$) stars \citep{1999ApJS..123....3L}.

\section{Correlation Between the Infrared Continuum and Mid-infrared Emission Lines}

Besides the low ionization mid-infrared [Ne~II] emission-line, the FIR is also a good indicator of  star formation, as it correlates very tightly  with the 1415 MHz radio luminosity, which is  thought to be produced by the same population of massive stars that heat and ionize H~II regions \citep{1992ARA&A..30..575C}.  There is a strong correlation between the FIR  and [Ne II] \citep{2002A&A...393..821S,2006ApJ...649...79S} which 
supports an scenario in which  the mid-infrared luminosity is dominated by the AGN, while  the far-infrared  luminosity is dominated by star formation \citep{2002A&A...393..821S,2006A&A...457L..17H}. 
\begin{figure}
\epsscale{.90}
\plotone{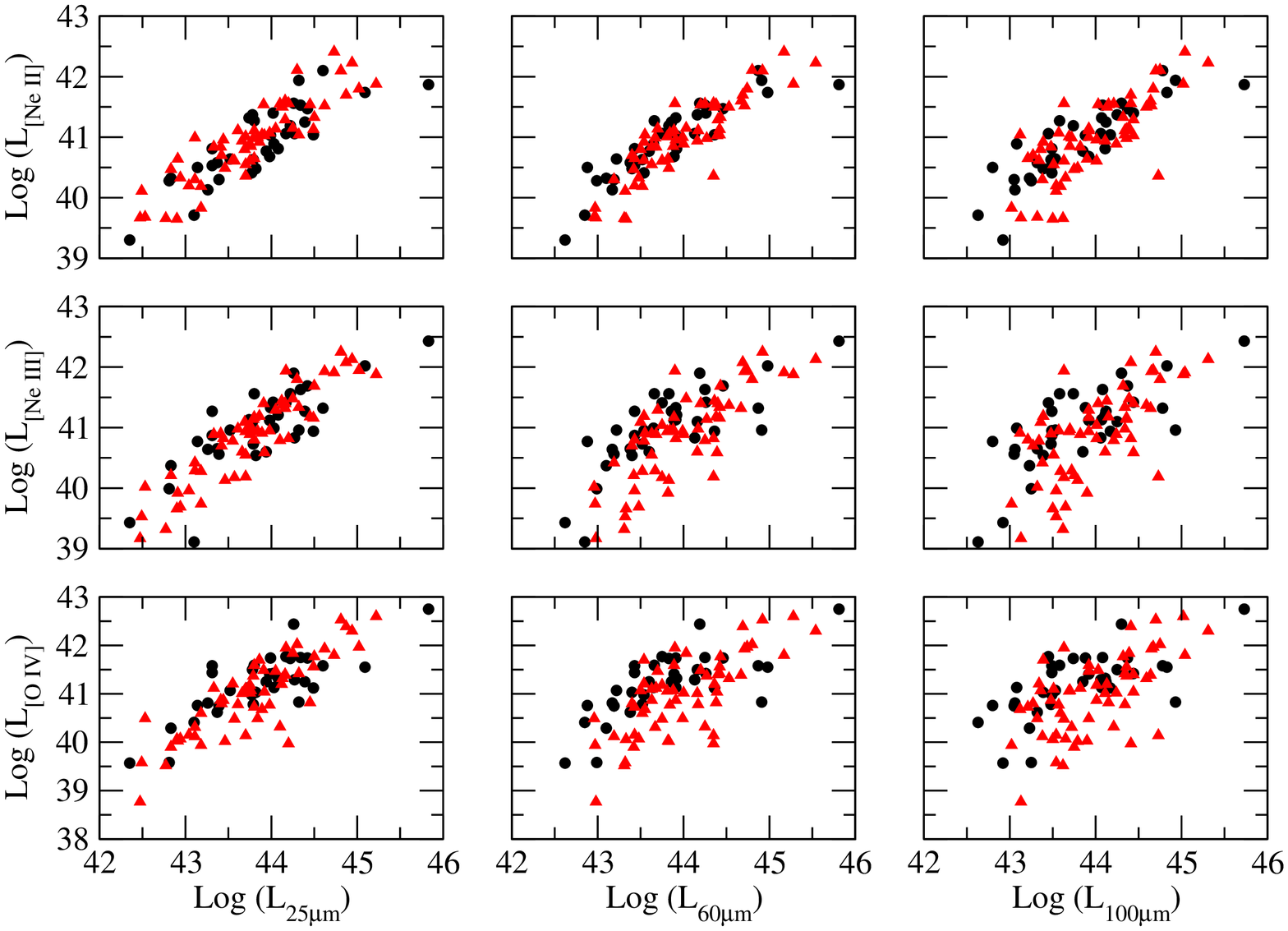}
    \caption{Correlation between the IR continuum   and mid-infrared emission line luminosities. The statistical analysis for the different  correlations  
between the mid-infrared emission lines and mid- and far-infrared continuum are presented in Table~\ref{corre1}. Symbols are identical to those in Figure~\ref{ratio1}.\label{ir_fig20}}
\end{figure}

 Figure~\ref{ir_fig20} shows the correlation between the IR continuum and mid-infrared emission lines in our sample. Table~\ref{corre1} shows the statistical analysis for the different correlations. We found that the [Ne~II] is well correlated with the continuum luminosity at 60$\micron$, in agreement with 
previous studies \citep[e.g.,][]{2002A&A...393..821S,2006ApJ...649...79S}. Compared with  the [Ne~II] correlation, both the  [Ne~III] and [O~IV] show larger scatter with respect to the IR continuum (see Table~\ref{corre1}). We have 
demonstrated  that [Ne~III] correlates with  [O~IV] suggesting [Ne~III] as a  good tracer of the AGN luminosity (see Section~4); the better correlation at shorter continuum 
wavelengths suggests a larger AGN contribution at those wavelengths in agreement with previous studies.

As we mentioned before, there is the technical difficulty in isolating the AGN continuum from the host galaxy emission \citep[e.g.,][]{2004A&A...418..465L}, specially 
in the larger field of view of {\it IRAS}. In order to estimate the star formation contribution in the mid- and far-infrared continuum  we used
  the correlations between the  [O~IV] and the mid- and far-infrared continuum in the  sources that show no  PAH features in their  spectra, as a template to estimate the contribution from  the  AGN. This contribution is inevitably mixed with some fraction of  star formation in   the host galaxy. By subtracting this contribution to the observed continuum luminosities we obtained the remaining 
fraction of star formation, e.g., a ``pure" stellar component, thus, a lower limit for the star formation contribution to the mid- and far-infrared continuum.

\begin{figure}
\epsscale{.90}
\plotone{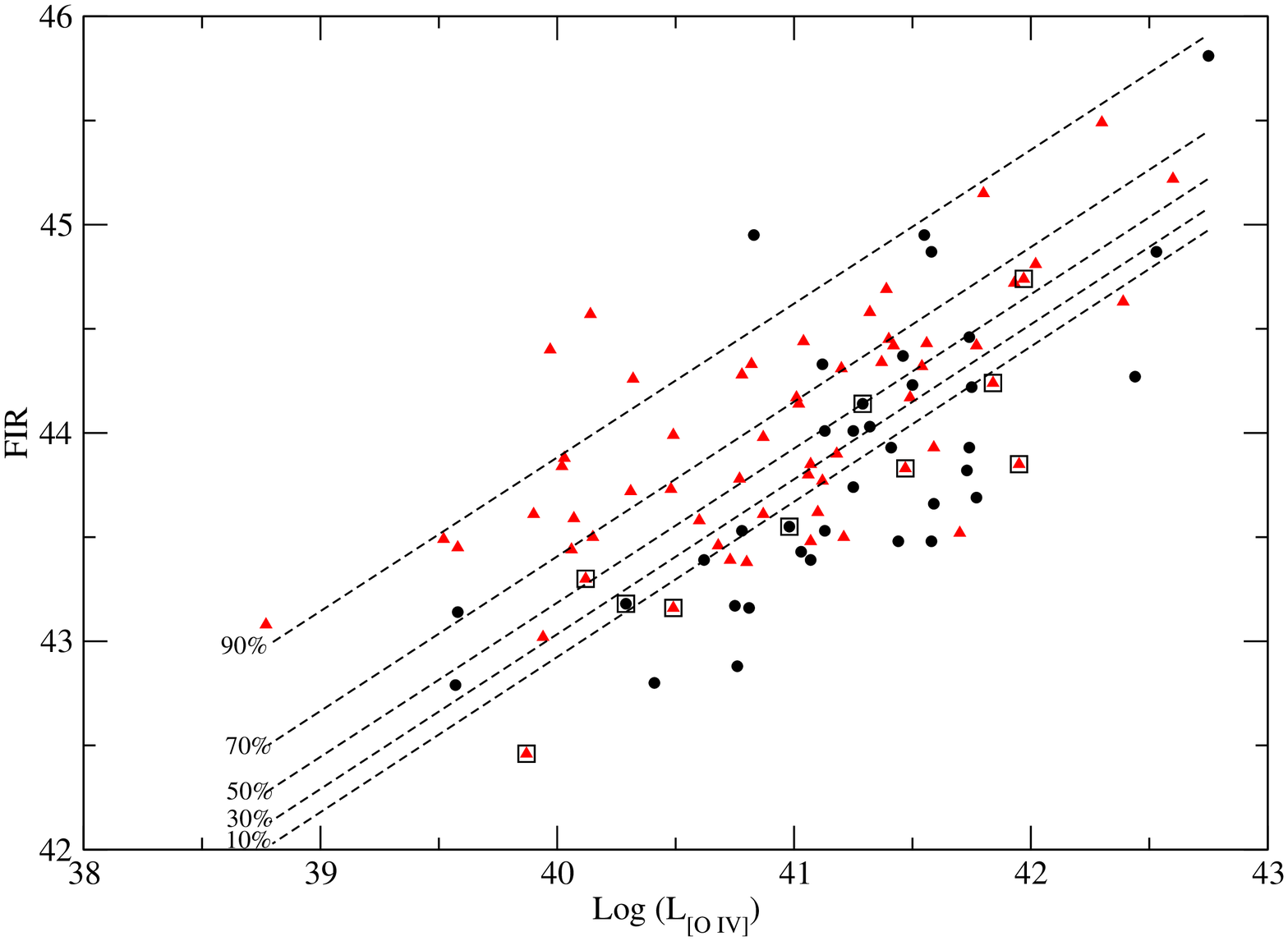}
    \caption{Correlation between the FIR luminosities and [O~IV]. The {\it dashed lines} represent the percentage of ``pure" stellar component  in the
FIR, based on the combined AGN+host galaxy stellar contribution. The squares represent the pure AGN sources, i.e., no detectable PAH features 
at 6.2$\micron$ and 11.2$\micron$. Symbols are identical to those in Figure~\ref{ratio1}.\label{ir_fig19}}
\end{figure}

In our sample,  we found the contribution of the star formation  that cannot be associated with the pure AGN sources to be:
$32\pm 2\%$, $45\pm 5\%$, $39 \pm 4\%$ and $42\pm 4\%$ for the luminosities at  25$\micron$, 60$\micron$, 100$\micron$ and FIR, respectively. These results suggest 
that   the far-infrared continuum contains a higher fraction from a stellar component than that found in the mid-infrared (25$\micron$).  Within  this  sample, 
Seyfert 1 galaxies exhibit  a narrower range of star formation contribution, $\sim 26 \pm 1\%$, to their mid- and far-infrared continuum luminosities, than 
that found for Seyfert 2 galaxies, $\sim 47 \pm 9 \%$. Figure~\ref{ir_fig19} shows the correlation between FIR and [O~IV] luminosities with the percentage of star formation contribution to the FIR luminosities indicated. The fact that most of the  pure AGN sources are in the lower part of the FIR-[O~IV] correlation, suggests that the fraction of 
 star formation in the host galaxy that is mixed  with the AGN contribution is minimal. Overall, these results are in good agreement from previous studies that used the infrared continuum to separate the relative contribution of star formation and nuclear activity \citep[e.g.,][]{2007ApJ...669..841S} and with the values derived from the [O~IV]-[Ne~II] correlation, which shows that Seyfert 2 galaxies have, on average, a stronger star formation contribution.

In order to assess the relative contribution from the AGN to the mid- and far-infrared continuum within Seyfert classification, we investigated the ratio of 
[O~IV] with the 25$\micron$, 60$\micron$ and FIR continuum luminosities. We found that Seyfert 1 and Seyfert 2 galaxies are statistically different in their 
AGN contribution to their 60$\micron$ and FIR luminosities with a probability of the null hypothesis of 0.2$\%$ and 0.4$\%$, respectively. On the other hand, 
the K-S test to the [O~IV]/$25\micron$ ratio shows that two samples drawn from the same population would differ this much only $\sim 36.7\%$ of the time.

\section{Physical Conditions in the [Ne~II] Emitting Region}

We have established that a fraction of  the [Ne~II] emission must come from the AGN, given the fact that  this line 
is present in the spectra of AGN that have no detected PAH features at 6.2$\micron$ and $11.5\micron$ (which have been shown to be tracers of star formation activity).
In order to investigate the physical conditions in the emission line regions, ionized by  the AGN, for  [\ion{Ne}{2}], [\ion{Ne}{3}], and [\ion{O}{4}] we generated
  a grid of dust-free, single-zone, constant-density models using  the photoionization code CLOUDY, version 07.02.01, last described by \cite{1998PASP..110..761F}. In this grid, 
hydrogen  density ($n_H$) and ionization parameter $U$ were  varied. The ionization parameter  $U$ is defined as \citep[see][]{2006agna.book.....O}:
\begin{equation}
U=\frac{1}{4\pi R^2cn_H}\int^\infty_{\nu_o}\frac{L_\nu}{h\nu}d\nu=\frac{Q(H)}{4\pi R^2cn_H},
\label{u}
\end{equation}
where R is the  distance to the cloud 
, c is the speed of light  and $Q(H)$ is the flux of ionizing photons.

We used a set of  roughly solar abundances \cite[e.g.,][]{1989AIPC..183....1G}. The logs of the abundances relative to H by number are: He: -1; C: -3.46; N: -3.92; O: -3.19;  Ne: -3.96;  Na: -5.69;  Mg: -4.48; Al: -5.53;
Si: -4.50; P: -6.43;  S: -4.82; Ar: -5.40;  Ca: -5.64;  Fe: -4.40 and  Ni: -5.75. We assumed a column density of $10^{21}{\rm cm^{-2}}$, which is  typical
 of the narrow line region  \citep[e.g.,][]{2000ApJ...531..278K}.

Assuming a distance of $R=130~{\rm pc}$ for the NLR (see M08), we overlaid  the observed mid-infrared emission line fluxes correlations (i.e., [Ne~II]-[Ne~III], [Ne~II]-[O~IV] and [Ne~III]-[O~IV])  from the ``pure" AGN sources with those obtained from the photoionization modeling. From this comparison  we derived the 
parameter space ($U,n_H$) required  to reproduce the observed relationships within their given dispersion. The predicted, intrinsic  fluxes from the photoionization models  are the line fluxes emitted at the ionized face of the slab of gas, 
used to model the NLR. Caution must be taken when comparing [Ne~II], [Ne~III] and [O~IV] given the fact that  the observed [Ne~II] fluxes can have contribution from both a star
 formation and  AGN. In this regard,  we are only interested in  the AGN component of [Ne~II], as deconvolved with the  method presented in this work. On the other hand, the observed [Ne~III] and [O~IV] in our sample have been assumed to represent the AGN power.

 From  the observed [Ne~III]/[Ne~II] ratios   we  obtained a range in   ionization parameter  $-4.00 < \log (U) < -3.50$ and for the hydrogen density $ 4.25~{\rm (cm^{-3})} < \log (n_H) < 5.50~{\rm (cm^{-3})}$. From the same set of models, we investigated the relationship between the AGN component of [Ne~II] and [O~IV] emission fluxes. From 
 the  observed range of [O~IV]/[Ne~II] ratios  we obtained  $-3.20 < \log (U) < -2.45$ and $ 3.25~{\rm (cm^{-3})} < \log (n_H) < 4.50~{\rm (cm^{-3})}$. Finally, we investigated the ratios  between the [O~IV]
 and [Ne~III]  emission fluxes.  We obtained a range for the ionization parameter,  $-3.20 < \log (U) < -1.65$, and for the  
 hydrogen density,  $ 2.00~{\rm (cm^{-3})} < \log (n_H) < 4.50~{\rm (cm^{-3})}$. This range in parameter space is the closest match to the one  found 
from the [O~IV]/[O~III] ratio (M08),  $-1.50 < \log (U) < -1.30$ and $ 2.0~{\rm (cm^{-3})} < \log (n_H) < 4.25~{\rm (cm^{-3})}$.

\begin{figure}
\epsscale{.90}
\plotone{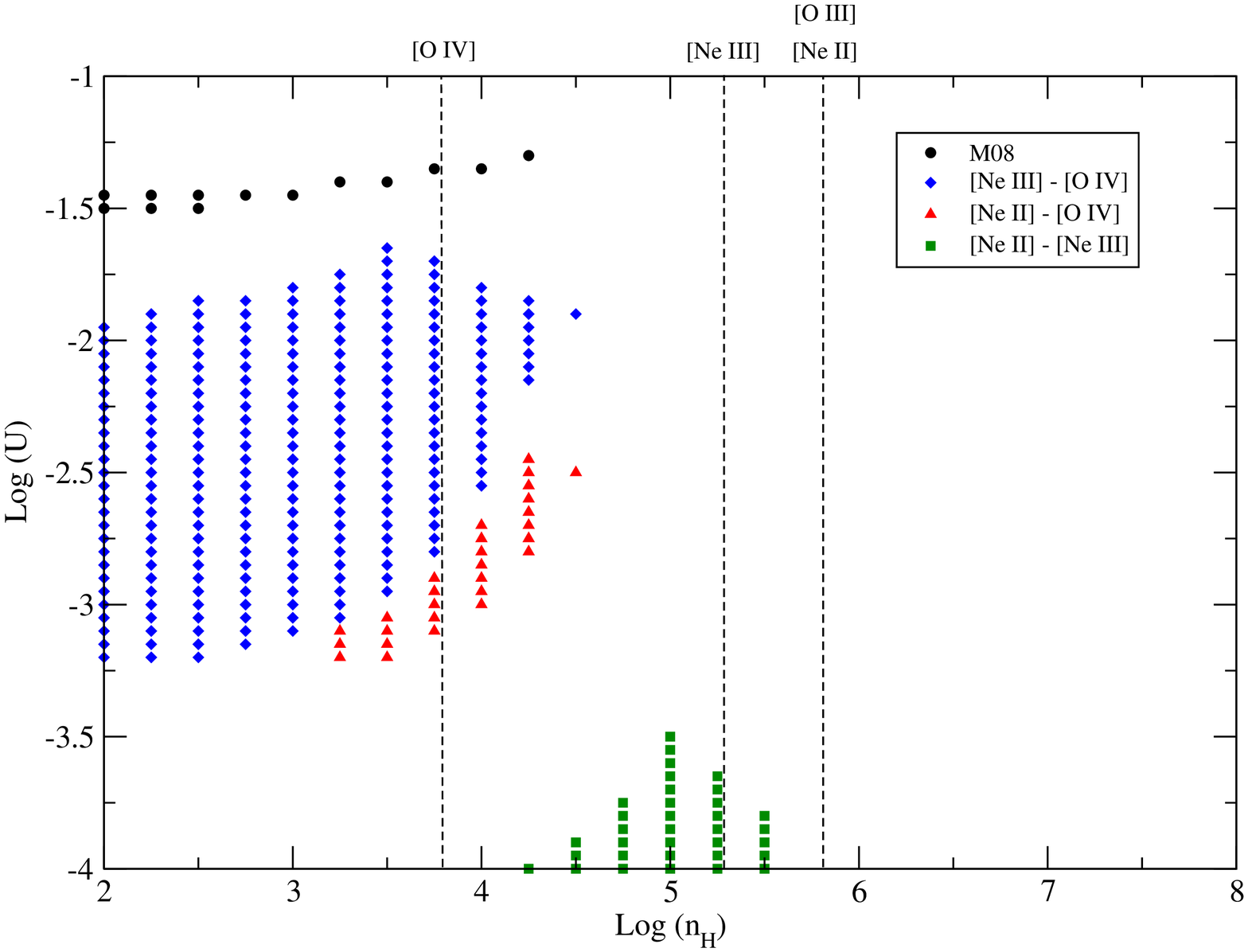}
    \caption{Comparison  between the parameter space required for the different mid-infrared line relationships. The dashed lines represent the critical 
 density for the different emission lines.\label{av_midinfrared_1}}
\end{figure}

In Figure~\ref{av_midinfrared_1} we show the allowed range in parameter space  from the different   emission line correlations and compared them with the 
parameter space obtained from the [O~IV]/[O~III] ratios (M08). Given 
the low ionization parameter obtained for [Ne~II], it is possible that the [Ne~II] that is  produced by the AGN 
   could originate in a more distant  region that [O~IV].  On the other hand, the [Ne~II]-[O~IV] relationship suggests a different [Ne~II] component at higher  ionization 
 and  lower densities than that  found from the [Ne~II]~-~[Ne~III] correlation,  indicating a  more closer and/or compact [Ne~II] emitting region. Another possibility is that [Ne~II] forms in regions that  are irradiated by the continuum filtered by ionized gas \citep{2008ApJ...679.1128K}. However for uncovered gas, there is only a
 small range in parameter space where the model successfully predicts the [Ne~II] emission associated with the AGN. 

\section{Conclusions}

We have investigated the ionization state of the emission-line gas  in Seyfert  galaxies  with the aim 
 of constraining the active galactic nuclei (AGN) and star formation contributions in the mid- and far-infrared spectra 
 for a sample of 103 Seyfert galaxies. We found the ratio  between the AGN power and star formation, as shown  by [O~IV]/[Ne~II], to be smaller in Seyfert 2 than 
in Seyfert 1 galaxies. In this regard, we also found a  correlation between [Ne~III] and [O~IV] versus  [Ne~II], 
with a clear separation between Seyfert groups. This separation suggests  that the [Ne~II] emission has two different contributions:  a component that could be associated 
 with  the AGN ionizing continuum and a component produced by photoionization from  star formation regions. The evidence for the former is   the presence of  [Ne~II] in Seyfert galaxies with no detectable PAH features at 6.2~$\mu$m and 11.5~$\mu$m. We also obtained a strong correlation between [Ne~III] and [O~IV] confirming that [Ne~III]
 can also be used to  estimate  the intrinsic power of the AGN.

We used the [O~IV] and [Ne~II] correlation  from the sources with no detectable PAH features as a template  to deconvolve the star formation and AGN contribution in the [Ne~II] emission. We found that Seyfert 2 galaxies, on average, 
have a relatively higher star formation contribution in their [Ne~II], by a factor of $\sim 1.5$, than Seyfert 1 galaxies, in agreement 
with other observations and theoretical work 
 on circumnuclear regions of AGN \citep[e.g.,][]{1985MNRAS.213..841T,1989ApJ...342..735H,1997ApJ...485..552M,2001ApJ...546..845G,2007ApJ...671..124D}. Using the stellar [Ne~II] luminosity as a SFR estimator we found that
  Seyfert 2 galaxies have similar  star formation rates
 in their spectra,  $8 \pm 2 M_\sun{\rm ~yr^{-1}}$ to that
found for  Seyfert 1 galaxies, $7\pm 2 M_\sun{\rm ~yr^{-1}}$. This result must be interpreted carefully given the fact that
  [Ne~II] emission provides an instantaneous measure of the star formation rate, independent 
of the previous star formation history. Thus, caution most be taken when comparing with other SFR indicators on galaxies with recent starbursts, but are no longer forming 
massive stars, in which case our star formation predictions may account only for a small fraction of a give SFR. Overall 
we found that $77\%$ of the Seyfert 2 galaxies in our sample show  some star formation contribution to their [Ne~II] observed luminosities 
while $56\%$ of the Seyfert 1 galaxies have some stellar component.

We  used the correlations between the mid- and far-infrared continuum  with the [O~IV]  luminosities in the non-PAH sources  to estimate
 the star formation contribution in the mid- and far-infrared continua. The resulting star formation contribution represents a lower limit  given the fact that 
some fraction of the star formation from the host galaxy is mixed with the AGN contribution in the mid- and far-infrared continuum in our sample of pure AGN sources. Averaged over all the sample we found the  contribution from star formation to the 25$\micron$, 60$\micron$, 100$\micron$ and FIR luminosities to be: $32\pm 2\%$, $45\pm 5\%$, $39 \pm 4\%$ and $42\pm 4\%$, respectively. We also found that Seyfert 1 galaxies exhibit  a narrower range of star formation contribution, $\sim 26 \pm 1\%$, to their mid- and far-infrared continuum luminosities, than that found for Seyfert 2 galaxies, $\sim 47 \pm 9 \%$.

 We also found a good correlation between [Ne~II] and mid- and far-infrared  luminosities, with the strongest  correlation 
for the [Ne~II]-60$\micron$. This result is in agreement 
with previous studies that link both the FIR and [Ne~II] emission with star formation. We  found a weaker correlation between [O~IV] and 
[Ne~III] versus the FIR luminosities. This result suggest that part of the FIR luminosity is reprocessed photons from the AGN, most likely  through  
reradiation  of  the AGN continuum by dust.   Assuming
 that the infrared continuum is dominated by emission from dust grains,  we found
 in our sample, that Seyfert 2 galaxies possess   cooler dust, with an average $\alpha_{25-60}=-1.5\pm0.1$,  than Seyfert 1 galaxies which have 
$\alpha_{25-60}=-0.8\pm0.1$.  In agreement, we also found that Seyfert 1 and Seyfert 2 galaxies are statistically different in terms 
of their  relative contribution from the AGN to the 60$\micron$ and FIR but have a similar contribution in the $25\micron$.

Summarizing, we found that Seyfert 1 and Seyfert 2 galaxies are statistically different in terms of the relative contribution from  the AGN and star formation
,as showed by the [O~IV]/[Ne~II] ratios. From this we found that Seyfert 2 galaxies have a higher star formation contribution, 
relative to the strength of the AGN, than that found in  Seyfert 1 galaxies but have  similar star formation rates. These results suggest that the differences between Seyfert galaxies cannot be  solely due to  viewing angle dependence.

\acknowledgments
We would like to thank our anonymous referee for her/his  suggestions  that improved the  paper. This research has made use of NASA's Astrophysics Data System. Also, this research has made use of the NASA/IPAC Extragalactic Database (NED) which is operated by the Jet Propulsion Laboratory, 
California Institute of Technology, under contract with the National Aeronautics and Space Administration. The IRS was a collaborative venture between 
Cornell University and Ball Aerospace Corporation funded by NASA through the Jet Propulsion Laboratory and Ames Research Center.
SMART was developed by the IRS Team at Cornell University and is available through the Spitzer Science Center at Caltech. Basic research in astronomy at the NRL is supported by 6.1 base funding.

\clearpage

\bibliographystyle{apj}

\bibliography{ms}

\begin{thebibliography}{63}
\expandafter\ifx\csname natexlab\endcsname\relax\def\natexlab#1{#1}\fi

\bibitem[{{Abel} \& {Satyapal}(2008)}]{2008ApJ...678..686A}
{Abel}, N.~P., \& {Satyapal}, S. 2008, \apj, 678, 686

\bibitem[{{Antonucci}(1993)}]{1993ARA&A..31..473A}
{Antonucci}, R. 1993, \araa, 31, 473

\bibitem[{{Antonucci} \& {Miller}(1985)}]{1985ApJ...297..621A}
{Antonucci}, R.~R.~J., \& {Miller}, J.~S. 1985, \apj, 297, 621

\bibitem[{{Armus} {et~al.}(2007){Armus}, {Charmandaris}, {Bernard-Salas},
  {Spoon}, {Marshall}, {Higdon}, {Desai}, {Teplitz}, {Hao}, {Devost}, {Brandl},
  {Wu}, {Sloan}, {Soifer}, {Houck}, \& {Herter}}]{2007ApJ...656..148A}
{Armus}, L., {Charmandaris}, V., {Bernard-Salas}, J., {Spoon}, H.~W.~W.,
  {Marshall}, J.~A., {Higdon}, S.~J.~U., {Desai}, V., {Teplitz}, H.~I., {Hao},
  L., {Devost}, D., {Brandl}, B.~R., {Wu}, Y., {Sloan}, G.~C., {Soifer}, B.~T.,
  {Houck}, J.~R., \& {Herter}, T.~L. 2007, \apj, 656, 148

\bibitem[{{Bevington} \& {Robinson}(2003)}]{2003drea.book.....B}
{Bevington}, P.~R., \& {Robinson}, D.~K. 2003, {Data reduction and error
  analysis for the physical sciences} (Data reduction and error analysis for
  the physical sciences, 3rd ed., by Philip R.~Bevington, and Keith
  D.~Robinson.~Boston, MA: McGraw-Hill, ISBN 0-07-247227-8, 2003.)

\bibitem[{{Buchanan} {et~al.}(2006){Buchanan}, {Gallimore}, {O'Dea}, {Baum},
  {Axon}, {Robinson}, {Elitzur}, \& {Elvis}}]{2006AJ....132..401B}
{Buchanan}, C.~L., {Gallimore}, J.~F., {O'Dea}, C.~P., {Baum}, S.~A., {Axon},
  D.~J., {Robinson}, A., {Elitzur}, M., \& {Elvis}, M. 2006, \aj, 132, 401

\bibitem[{{Calzetti} {et~al.}(2005){Calzetti}, {Kennicutt}, {Bianchi},
  {Thilker}, {Dale}, {Engelbracht}, {Leitherer}, {Meyer}, {Sosey}, {Mutchler},
  {Regan}, {Thornley}, {Armus}, {Bendo}, {Boissier}, {Boselli}, {Draine},
  {Gordon}, {Helou}, {Hollenbach}, {Kewley}, {Madore}, {Martin}, {Murphy},
  {Rieke}, {Rieke}, {Roussel}, {Sheth}, {Smith}, {Walter}, {White}, {Yi},
  {Scoville}, {Polletta}, \& {Lindler}}]{2005ApJ...633..871C}
{Calzetti}, D., {Kennicutt}, Jr., R.~C., {Bianchi}, L., {Thilker}, D.~A.,
  {Dale}, D.~A., {Engelbracht}, C.~W., {Leitherer}, C., {Meyer}, M.~J.,
  {Sosey}, M.~L., {Mutchler}, M., {Regan}, M.~W., {Thornley}, M.~D., {Armus},
  L., {Bendo}, G.~J., {Boissier}, S., {Boselli}, A., {Draine}, B.~T., {Gordon},
  K.~D., {Helou}, G., {Hollenbach}, D.~J., {Kewley}, L., {Madore}, B.~F.,
  {Martin}, D.~C., {Murphy}, E.~J., {Rieke}, G.~H., {Rieke}, M.~J., {Roussel},
  H., {Sheth}, K., {Smith}, J.~D., {Walter}, F., {White}, B.~A., {Yi}, S.,
  {Scoville}, N.~Z., {Polletta}, M., \& {Lindler}, D. 2005, \apj, 633, 871

\bibitem[{{Cid Fernandes} {et~al.}(2004){Cid Fernandes}, {Gu}, {Melnick},
  {Terlevich}, {Terlevich}, {Kunth}, {Rodrigues Lacerda}, \&
  {Joguet}}]{2004MNRAS.355..273C}
{Cid Fernandes}, R., {Gu}, Q., {Melnick}, J., {Terlevich}, E., {Terlevich}, R.,
  {Kunth}, D., {Rodrigues Lacerda}, R., \& {Joguet}, B. 2004, \mnras, 355, 273

\bibitem[{{Cid Fernandes} {et~al.}(2001){Cid Fernandes}, {Heckman}, {Schmitt},
  {Delgado}, \& {Storchi-Bergmann}}]{2001ApJ...558...81C}
{Cid Fernandes}, R., {Heckman}, T., {Schmitt}, H., {Delgado}, R.~M.~G., \&
  {Storchi-Bergmann}, T. 2001, \apj, 558, 81

\bibitem[{{Cid Fernandes} \& {Terlevich}(1995)}]{1995MNRAS.272..423C}
{Cid Fernandes}, R.~J., \& {Terlevich}, R. 1995, \mnras, 272, 423

\bibitem[{{Clavel} {et~al.}(2000){Clavel}, {Schulz}, {Altieri}, {Barr},
  {Claes}, {Heras}, {Leech}, {Metcalfe}, \& {Salama}}]{2000A&A...357..839C}
{Clavel}, J., {Schulz}, B., {Altieri}, B., {Barr}, P., {Claes}, P., {Heras},
  A., {Leech}, K., {Metcalfe}, L., \& {Salama}, A. 2000, \aap, 357, 839

\bibitem[{{Condon}(1992)}]{1992ARA&A..30..575C}
{Condon}, J.~J. 1992, \araa, 30, 575

\bibitem[{{Dale} {et~al.}(2006){Dale}, {Smith}, {Armus}, {Buckalew}, {Helou},
  {Kennicutt}, {Moustakas}, {Roussel}, {Sheth}, {Bendo}, {Calzetti}, {Draine},
  {Engelbracht}, {Gordon}, {Hollenbach}, {Jarrett}, {Kewley}, {Leitherer},
  {Li}, {Malhotra}, {Murphy}, \& {Walter}}]{2006ApJ...646..161D}
{Dale}, D.~A., {Smith}, J.~D.~T., {Armus}, L., {Buckalew}, B.~A., {Helou}, G.,
  {Kennicutt}, Jr., R.~C., {Moustakas}, J., {Roussel}, H., {Sheth}, K.,
  {Bendo}, G.~J., {Calzetti}, D., {Draine}, B.~T., {Engelbracht}, C.~W.,
  {Gordon}, K.~D., {Hollenbach}, D.~J., {Jarrett}, T.~H., {Kewley}, L.~J.,
  {Leitherer}, C., {Li}, A., {Malhotra}, S., {Murphy}, E.~J., \& {Walter}, F.
  2006, \apj, 646, 161

\bibitem[{{Davies} {et~al.}(2007){Davies}, {Mueller S{\'a}nchez}, {Genzel},
  {Tacconi}, {Hicks}, {Friedrich}, \& {Sternberg}}]{2007ApJ...671.1388D}
{Davies}, R.~I., {Mueller S{\'a}nchez}, F., {Genzel}, R., {Tacconi}, L.~J.,
  {Hicks}, E.~K.~S., {Friedrich}, S., \& {Sternberg}, A. 2007, \apj, 671, 1388

\bibitem[{{Deo} {et~al.}(2007){Deo}, {Crenshaw}, {Kraemer}, {Dietrich},
  {Elitzur}, {Teplitz}, \& {Turner}}]{2007ApJ...671..124D}
{Deo}, R.~P., {Crenshaw}, D.~M., {Kraemer}, S.~B., {Dietrich}, M., {Elitzur},
  M., {Teplitz}, H., \& {Turner}, T.~J. 2007, \apj, 671, 124

\bibitem[{{Ferland} {et~al.}(1998){Ferland}, {Korista}, {Verner}, {Ferguson},
  {Kingdon}, \& {Verner}}]{1998PASP..110..761F}
{Ferland}, G.~J., {Korista}, K.~T., {Verner}, D.~A., {Ferguson}, J.~W.,
  {Kingdon}, J.~B., \& {Verner}, E.~M. 1998, \pasp, 110, 761

\bibitem[{{F{\"o}rster Schreiber} {et~al.}(2004){F{\"o}rster Schreiber},
  {Roussel}, {Sauvage}, \& {Charmandaris}}]{2004A&A...419..501F}
{F{\"o}rster Schreiber}, N.~M., {Roussel}, H., {Sauvage}, M., \&
  {Charmandaris}, V. 2004, \aap, 419, 501

\bibitem[{{Genzel} {et~al.}(1998){Genzel}, {Lutz}, {Sturm}, {Egami}, {Kunze},
  {Moorwood}, {Rigopoulou}, {Spoon}, {Sternberg}, {Tacconi-Garman}, {Tacconi},
  \& {Thatte}}]{1998ApJ...498..579G}
{Genzel}, R., {Lutz}, D., {Sturm}, E., {Egami}, E., {Kunze}, D., {Moorwood},
  A.~F.~M., {Rigopoulou}, D., {Spoon}, H.~W.~W., {Sternberg}, A.,
  {Tacconi-Garman}, L.~E., {Tacconi}, L., \& {Thatte}, N. 1998, \apj, 498, 579

\bibitem[{{Gonz{\'a}lez Delgado} {et~al.}(2001){Gonz{\'a}lez Delgado},
  {Heckman}, \& {Leitherer}}]{2001ApJ...546..845G}
{Gonz{\'a}lez Delgado}, R.~M., {Heckman}, T., \& {Leitherer}, C. 2001, \apj,
  546, 845

\bibitem[{{Gorjian} {et~al.}(2007){Gorjian}, {Cleary}, {Werner}, \&
  {Lawrence}}]{2007ApJ...655L..73G}
{Gorjian}, V., {Cleary}, K., {Werner}, M.~W., \& {Lawrence}, C.~R. 2007, \apjl,
  655, L73

\bibitem[{{Grevesse} \& {Anders}(1989)}]{1989AIPC..183....1G}
{Grevesse}, N., \& {Anders}, E. 1989, in American Institute of Physics
  Conference Series, Vol. 183, Cosmic Abundances of Matter, ed. C.~J.
  {Waddington}, 1--8

\bibitem[{{Gu} {et~al.}(2006){Gu}, {Melnick}, {Fernandes}, {Kunth},
  {Terlevich}, \& {Terlevich}}]{2006MNRAS.366..480G}
{Gu}, Q., {Melnick}, J., {Fernandes}, R.~C., {Kunth}, D., {Terlevich}, E., \&
  {Terlevich}, R. 2006, \mnras, 366, 480

\bibitem[{{Heckman} {et~al.}(1989){Heckman}, {Blitz}, {Wilson}, {Armus}, \&
  {Miley}}]{1989ApJ...342..735H}
{Heckman}, T.~M., {Blitz}, L., {Wilson}, A.~S., {Armus}, L., \& {Miley}, G.~K.
  1989, \apj, 342, 735

\bibitem[{{Heisler} {et~al.}(1997){Heisler}, {Lumsden}, \&
  {Bailey}}]{1997Natur.385..700H}
{Heisler}, C.~A., {Lumsden}, S.~L., \& {Bailey}, J.~A. 1997, \nat, 385, 700

\bibitem[{{Ho} {et~al.}(2003){Ho}, {Filippenko}, \&
  {Sargent}}]{2003ApJ...583..159H}
{Ho}, L.~C., {Filippenko}, A.~V., \& {Sargent}, W.~L.~W. 2003, \apj, 583, 159

\bibitem[{{Ho} \& {Keto}(2007)}]{2007ApJ...658..314H}
{Ho}, L.~C., \& {Keto}, E. 2007, \apj, 658, 314

\bibitem[{{Horst} {et~al.}(2006){Horst}, {Smette}, {Gandhi}, \&
  {Duschl}}]{2006A&A...457L..17H}
{Horst}, H., {Smette}, A., {Gandhi}, P., \& {Duschl}, W.~J. 2006, \aap, 457,
  L17

\bibitem[{{Houck} {et~al.}(2004){Houck}, {Roellig}, {van Cleve}, {Forrest},
  {Herter}, {Lawrence}, {Matthews}, {Reitsema}, {Soifer}, {Watson}, {Weedman},
  {Huisjen}, {Troeltzsch}, {Barry}, {Bernard-Salas}, {Blacken}, {Brandl},
  {Charmandaris}, {Devost}, {Gull}, {Hall}, {Henderson}, {Higdon}, {Pirger},
  {Schoenwald}, {Sloan}, {Uchida}, {Appleton}, {Armus}, {Burgdorf},
  {Fajardo-Acosta}, {Grillmair}, {Ingalls}, {Morris}, \&
  {Teplitz}}]{2004ApJS..154...18H}
{Houck}, J.~R., {Roellig}, T.~L., {van Cleve}, J., {Forrest}, W.~J., {Herter},
  T., {Lawrence}, C.~R., {Matthews}, K., {Reitsema}, H.~J., {Soifer}, B.~T.,
  {Watson}, D.~M., {Weedman}, D., {Huisjen}, M., {Troeltzsch}, J., {Barry},
  D.~J., {Bernard-Salas}, J., {Blacken}, C.~E., {Brandl}, B.~R.,
  {Charmandaris}, V., {Devost}, D., {Gull}, G.~E., {Hall}, P., {Henderson},
  C.~P., {Higdon}, S.~J.~U., {Pirger}, B.~E., {Schoenwald}, J., {Sloan}, G.~C.,
  {Uchida}, K.~I., {Appleton}, P.~N., {Armus}, L., {Burgdorf}, M.~J.,
  {Fajardo-Acosta}, S.~B., {Grillmair}, C.~J., {Ingalls}, J.~G., {Morris},
  P.~W., \& {Teplitz}, H.~I. 2004, \apjs, 154, 18

\bibitem[{{Kennicutt}(1998)}]{1998ARA&A..36..189K}
{Kennicutt}, Jr., R.~C. 1998, \araa, 36, 189

\bibitem[{{Khachikian} \& {Weedman}(1974)}]{1974ApJ...192..581K}
{Khachikian}, E.~Y., \& {Weedman}, D.~W. 1974, \apj, 192, 581

\bibitem[{{Kraemer} {et~al.}(2000){Kraemer}, {Crenshaw}, {Hutchings}, {Gull},
  {Kaiser}, {Nelson}, \& {Weistrop}}]{2000ApJ...531..278K}
{Kraemer}, S.~B., {Crenshaw}, D.~M., {Hutchings}, J.~B., {Gull}, T.~R.,
  {Kaiser}, M.~E., {Nelson}, C.~H., \& {Weistrop}, D. 2000, \apj, 531, 278

\bibitem[{{Kraemer} {et~al.}(2008){Kraemer}, {Schmitt}, \&
  {Crenshaw}}]{2008ApJ...679.1128K}
{Kraemer}, S.~B., {Schmitt}, H.~R., \& {Crenshaw}, D.~M. 2008, \apj, 679, 1128

\bibitem[{{Leitherer} {et~al.}(1999){Leitherer}, {Schaerer}, {Goldader},
  {Delgado}, {Robert}, {Kune}, {de Mello}, {Devost}, \&
  {Heckman}}]{1999ApJS..123....3L}
{Leitherer}, C., {Schaerer}, D., {Goldader}, J.~D., {Delgado}, R.~M.~G.,
  {Robert}, C., {Kune}, D.~F., {de Mello}, D.~F., {Devost}, D., \& {Heckman},
  T.~M. 1999, \apjs, 123, 3

\bibitem[{{Low} {et~al.}(1984){Low}, {Young}, {Beintema}, {Gautier},
  {Beichman}, {Aumann}, {Gillett}, {Neugebauer}, {Boggess}, \&
  {Emerson}}]{1984ApJ...278L..19L}
{Low}, F.~J., {Young}, E., {Beintema}, D.~A., {Gautier}, T.~N., {Beichman},
  C.~A., {Aumann}, H.~H., {Gillett}, F.~C., {Neugebauer}, G., {Boggess}, N., \&
  {Emerson}, J.~P. 1984, \apjl, 278, L19

\bibitem[{{Lutz} {et~al.}(1996){Lutz}, {Genzel}, {Sternberg}, {Netzer},
  {Kunze}, {Rigopoulou}, {Sturm}, {Egami}, {Feuchtgruber}, {Moorwood}, \& {de
  Graauw}}]{1996A&A...315L.137L}
{Lutz}, D., {Genzel}, R., {Sternberg}, A., {Netzer}, H., {Kunze}, D.,
  {Rigopoulou}, D., {Sturm}, E., {Egami}, E., {Feuchtgruber}, H., {Moorwood},
  A.~F.~M., \& {de Graauw}, T. 1996, \aap, 315, L137

\bibitem[{{Lutz} {et~al.}(2004){Lutz}, {Maiolino}, {Spoon}, \&
  {Moorwood}}]{2004A&A...418..465L}
{Lutz}, D., {Maiolino}, R., {Spoon}, H.~W.~W., \& {Moorwood}, A.~F.~M. 2004,
  \aap, 418, 465

\bibitem[{{Maiolino} {et~al.}(1998){Maiolino}, {Krabbe}, {Thatte}, \&
  {Genzel}}]{1998ApJ...493..650M}
{Maiolino}, R., {Krabbe}, A., {Thatte}, N., \& {Genzel}, R. 1998, \apj, 493,
  650

\bibitem[{{Maiolino} {et~al.}(1997){Maiolino}, {Ruiz}, {Rieke}, \&
  {Papadopoulos}}]{1997ApJ...485..552M}
{Maiolino}, R., {Ruiz}, M., {Rieke}, G.~H., \& {Papadopoulos}, P. 1997, \apj,
  485, 552

\bibitem[{{Markwardt} {et~al.}(2005){Markwardt}, {Tueller}, {Skinner},
  {Gehrels}, {Barthelmy}, \& {Mushotzky}}]{2005ApJ...633L..77M}
{Markwardt}, C.~B., {Tueller}, J., {Skinner}, G.~K., {Gehrels}, N.,
  {Barthelmy}, S.~D., \& {Mushotzky}, R.~F. 2005, \apjl, 633, L77

\bibitem[{{Mel{\'e}ndez} {et~al.}(2008){Mel{\'e}ndez}, {Kraemer}, {Armentrout},
  {Deo}, {Crenshaw}, {Schmitt}, {Mushotzky}, {Tueller}, {Markwardt}, \&
  {Winter}}]{2008arXiv0804.1147M}
{Mel{\'e}ndez}, M., {Kraemer}, S.~B., {Armentrout}, B.~K., {Deo}, R.~P.,
  {Crenshaw}, D.~M., {Schmitt}, H.~R., {Mushotzky}, R.~F., {Tueller}, J.,
  {Markwardt}, C.~B., \& {Winter}, L. 2008, ArXiv e-prints, 804

\bibitem[{{Moshir} {et~al.}(1990){Moshir}, {Kopan}, {Conrow}, {McCallon},
  {Hacking}, {Gregorich}, {Rohrbach}, {Melnyk}, {Rice}, {Fullmer}, {White}, \&
  {Chester}}]{1990BAAS...22Q1325M}
{Moshir}, M., {Kopan}, G., {Conrow}, T., {McCallon}, H., {Hacking}, P.,
  {Gregorich}, D., {Rohrbach}, G., {Melnyk}, M., {Rice}, W., {Fullmer}, L.,
  {White}, J., \& {Chester}, T. 1990, in Bulletin of the American Astronomical
  Society, Vol.~22, Bulletin of the American Astronomical Society, 1325--+

\bibitem[{{Neugebauer} {et~al.}(1984){Neugebauer}, {Habing}, {van Duinen},
  {Aumann}, {Baud}, {Beichman}, {Beintema}, {Boggess}, {Clegg}, {de Jong},
  {Emerson}, {Gautier}, {Gillett}, {Harris}, {Hauser}, {Houck}, {Jennings},
  {Low}, {Marsden}, {Miley}, {Olnon}, {Pottasch}, {Raimond}, {Rowan-Robinson},
  {Soifer}, {Walker}, {Wesselius}, \& {Young}}]{1984ApJ...278L...1N}
{Neugebauer}, G., {Habing}, H.~J., {van Duinen}, R., {Aumann}, H.~H., {Baud},
  B., {Beichman}, C.~A., {Beintema}, D.~A., {Boggess}, N., {Clegg}, P.~E., {de
  Jong}, T., {Emerson}, J.~P., {Gautier}, T.~N., {Gillett}, F.~C., {Harris},
  S., {Hauser}, M.~G., {Houck}, J.~R., {Jennings}, R.~E., {Low}, F.~J.,
  {Marsden}, P.~L., {Miley}, G., {Olnon}, F.~M., {Pottasch}, S.~R., {Raimond},
  E., {Rowan-Robinson}, M., {Soifer}, B.~T., {Walker}, R.~G., {Wesselius},
  P.~R., \& {Young}, E. 1984, \apjl, 278, L1

\bibitem[{{Osterbrock} \& {Ferland}(2006)}]{2006agna.book.....O}
{Osterbrock}, D.~E., \& {Ferland}, G.~J. 2006, {Astrophysics of gaseous nebulae
  and active galactic nuclei} (Astrophysics of gaseous nebulae and active
  galactic nuclei, 2nd.~ed.~by D.E.~Osterbrock and G.J.~Ferland.~Sausalito, CA:
  University Science Books, 2006)

\bibitem[{{Perez-Olea} \& {Colina}(1996)}]{1996ApJ...468..191P}
{Perez-Olea}, D.~E., \& {Colina}, L. 1996, \apj, 468, 191

\bibitem[{{Peterson} {et~al.}(2004){Peterson}, {Ferrarese}, {Gilbert}, {Kaspi},
  {Malkan}, {Maoz}, {Merritt}, {Netzer}, {Onken}, {Pogge}, {Vestergaard}, \&
  {Wandel}}]{2004ApJ...613..682P}
{Peterson}, B.~M., {Ferrarese}, L., {Gilbert}, K.~M., {Kaspi}, S., {Malkan},
  M.~A., {Maoz}, D., {Merritt}, D., {Netzer}, H., {Onken}, C.~A., {Pogge},
  R.~W., {Vestergaard}, M., \& {Wandel}, A. 2004, \apj, 613, 682

\bibitem[{{Peterson} \& {Wandel}(2000)}]{2000ApJ...540L..13P}
{Peterson}, B.~M., \& {Wandel}, A. 2000, \apjl, 540, L13

\bibitem[{{Pier} \& {Krolik}(1992)}]{1992ApJ...401...99P}
{Pier}, E.~A., \& {Krolik}, J.~H. 1992, \apj, 401, 99

\bibitem[{{Press} {et~al.}(1992){Press}, {Teukolsky}, {Vetterling}, \&
  {Flannery}}]{1992nrfa.book.....P}
{Press}, W.~H., {Teukolsky}, S.~A., {Vetterling}, W.~T., \& {Flannery}, B.~P.
  1992, {Numerical recipes in FORTRAN. The art of scientific computing}
  (Cambridge: University Press, |c1992, 2nd ed.)

\bibitem[{{Rees}(1984)}]{1984ARA&A..22..471R}
{Rees}, M.~J. 1984, \araa, 22, 471

\bibitem[{{Sanders} {et~al.}(2003){Sanders}, {Mazzarella}, {Kim}, {Surace}, \&
  {Soifer}}]{2003AJ....126.1607S}
{Sanders}, D.~B., {Mazzarella}, J.~M., {Kim}, D.-C., {Surace}, J.~A., \&
  {Soifer}, B.~T. 2003, \aj, 126, 1607

\bibitem[{{Sanders} \& {Mirabel}(1996)}]{1996ARA&A..34..749S}
{Sanders}, D.~B., \& {Mirabel}, I.~F. 1996, \araa, 34, 749

\bibitem[{{Sanders} {et~al.}(1988){Sanders}, {Soifer}, {Elias}, {Neugebauer},
  \& {Matthews}}]{1988ApJ...328L..35S}
{Sanders}, D.~B., {Soifer}, B.~T., {Elias}, J.~H., {Neugebauer}, G., \&
  {Matthews}, K. 1988, \apjl, 328, L35

\bibitem[{{Schmitt} {et~al.}(2001){Schmitt}, {Antonucci}, {Ulvestad}, {Kinney},
  {Clarke}, \& {Pringle}}]{2001ApJ...555..663S}
{Schmitt}, H.~R., {Antonucci}, R.~R.~J., {Ulvestad}, J.~S., {Kinney}, A.~L.,
  {Clarke}, C.~J., \& {Pringle}, J.~E. 2001, \apj, 555, 663

\bibitem[{{Schmitt} {et~al.}(1999){Schmitt}, {Storchi-Bergmann}, \&
  {Fernandes}}]{1999MNRAS.303..173S}
{Schmitt}, H.~R., {Storchi-Bergmann}, T., \& {Fernandes}, R.~C. 1999, \mnras,
  303, 173

\bibitem[{{Schweitzer} {et~al.}(2006){Schweitzer}, {Lutz}, {Sturm}, {Contursi},
  {Tacconi}, {Lehnert}, {Dasyra}, {Genzel}, {Veilleux}, {Rupke}, {Kim},
  {Baker}, {Netzer}, {Sternberg}, {Mazzarella}, \&
  {Lord}}]{2006ApJ...649...79S}
{Schweitzer}, M., {Lutz}, D., {Sturm}, E., {Contursi}, A., {Tacconi}, L.~J.,
  {Lehnert}, M.~D., {Dasyra}, K.~M., {Genzel}, R., {Veilleux}, S., {Rupke}, D.,
  {Kim}, D.-C., {Baker}, A.~J., {Netzer}, H., {Sternberg}, A., {Mazzarella},
  J., \& {Lord}, S. 2006, \apj, 649, 79

\bibitem[{{Shi} {et~al.}(2007){Shi}, {Ogle}, {Rieke}, {Antonucci}, {Hines},
  {Smith}, {Low}, {Bouwman}, \& {Willmer}}]{2007ApJ...669..841S}
{Shi}, Y., {Ogle}, P., {Rieke}, G.~H., {Antonucci}, R., {Hines}, D.~C.,
  {Smith}, P.~S., {Low}, F.~J., {Bouwman}, J., \& {Willmer}, C. 2007, \apj,
  669, 841

\bibitem[{{Soifer} {et~al.}(1989){Soifer}, {Boehmer}, {Neugebauer}, \&
  {Sanders}}]{1989AJ.....98..766S}
{Soifer}, B.~T., {Boehmer}, L., {Neugebauer}, G., \& {Sanders}, D.~B. 1989,
  \aj, 98, 766

\bibitem[{{Spinoglio} {et~al.}(1995){Spinoglio}, {Malkan}, {Rush}, {Carrasco},
  \& {Recillas-Cruz}}]{1995ApJ...453..616S}
{Spinoglio}, L., {Malkan}, M.~A., {Rush}, B., {Carrasco}, L., \&
  {Recillas-Cruz}, E. 1995, \apj, 453, 616

\bibitem[{{Sturm} {et~al.}(2002){Sturm}, {Lutz}, {Verma}, {Netzer},
  {Sternberg}, {Moorwood}, {Oliva}, \& {Genzel}}]{2002A&A...393..821S}
{Sturm}, E., {Lutz}, D., {Verma}, A., {Netzer}, H., {Sternberg}, A.,
  {Moorwood}, A.~F.~M., {Oliva}, E., \& {Genzel}, R. 2002, \aap, 393, 821

\bibitem[{{Terlevich} {et~al.}(1990){Terlevich}, {Diaz}, \&
  {Terlevich}}]{1990MNRAS.242..271T}
{Terlevich}, E., {Diaz}, A.~I., \& {Terlevich}, R. 1990, \mnras, 242, 271

\bibitem[{{Terlevich} \& {Melnick}(1985)}]{1985MNRAS.213..841T}
{Terlevich}, R., \& {Melnick}, J. 1985, \mnras, 213, 841

\bibitem[{{Tommasin} {et~al.}(2008){Tommasin}, {Spinoglio}, {Malkan}, {Smith},
  {Gonz{\'a}lez-Alfonso}, \& {Charmandaris}}]{2007arXiv0710.4448T}
{Tommasin}, S., {Spinoglio}, L., {Malkan}, M.~A., {Smith}, H.,
  {Gonz{\'a}lez-Alfonso}, E., \& {Charmandaris}, V. 2008, \apj, 676, 836

\bibitem[{{Weedman} {et~al.}(2005){Weedman}, {Hao}, {Higdon}, {Devost}, {Wu},
  {Charmandaris}, {Brandl}, {Bass}, \& {Houck}}]{2005ApJ...633..706W}
{Weedman}, D.~W., {Hao}, L., {Higdon}, S.~J.~U., {Devost}, D., {Wu}, Y.,
  {Charmandaris}, V., {Brandl}, B., {Bass}, E., \& {Houck}, J.~R. 2005, \apj,
  633, 706

\end{thebibliography}
\clearpage

\clearpage

\begin{deluxetable}{lllllllllllll}
\tablecolumns{13}
\rotate
\tabletypesize{\scriptsize}
\tablewidth{0pt}
\tablecaption{The Infrared Sample of Seyfert Galaxies}
\tablehead{ 
\colhead{Name} & \colhead{Type} &\colhead{$z$}&\colhead{$\log {\rm L_{25\mu m}}$}&\colhead{$\log {\rm L_{60\mu m}}$}
&\colhead{$\log {\rm L_{FIR}}$}&\colhead{$\log {\rm L_{[Ne~II]}}$} & \colhead{$\log {\rm L_{[Ne~III]}}$} &
 \colhead{$\log {\rm L_{[O~IV]}}$} & \colhead{SC (\%)}& \colhead{SFR}& \colhead{Aperture} & \colhead{Reference} \\
&&& \multicolumn{6}{c}{(ergs ${\rm s^{-1}}$)}&&$({\rm M_\sun~yr^{-1}})$&(kpc)\\
}

\startdata

3C120	&	1	&	0.033010	&	44.26	&	44.19	&	44.27	&	41.56	&	41.90	&	42.44	&	0	&	0	&	3.23	&	1	\\
3C273	&	1	&	0.158339	&	45.83	&	45.81	&	45.81	&	41.87	&	42.43	&	42.75	&	0	&	0	&	16.45	&	1	\\
3C382	&	1	&	0.057870	&	43.94	&	43.60	&	43.74	&	40.77	&	40.60	&	41.25	&	0	&	0	&	5.73	&	1	\\
3C390.3	&	1	&	0.056100	&	44.39	&	43.86	&	44.01	&	41.25	&	41.27	&	41.25	&	56.11	&	5.77	&	5.55	&	1	\\
3C452	&	2	&	0.081100	&	\nodata	&	\nodata	&	\nodata	&	41.51	&	41.42	&	41.22	&	77.13	&	14.39	&	8.12	&	1	\\
CGCG381-051	&	2	&	0.030668	&	44.10	&	44.25	&	44.26	&	41.55	&	40.79	&	40.32	&	96.24	&	19.61	&	2.36	&	2	\\
Circinus	&	2	&	0.001448	&	43.57	&	43.75	&	43.73	&	40.61	&	40.18	&	40.48	&	55.00	&	1.28	&	0.14	&	4	\\
Cen A           &       2       &       0.001825        &       43.17   &       43.76   &       43.82   &       40.20   &       40.08   &       39.93   &       59.27
&       0.54    &       0.18    &       5       \\
ESO012-G021	&	1	&	0.030021	&	43.78	&	44.16	&	44.23	&	41.37	&	41.10	&	41.50	&	47.42	&	6.50	&	2.93	&	3	\\
ESO103-G035	&	1	&	0.013286	&	44.03	&	43.65	&	43.53	&	40.89	&	40.99	&	41.13	&	19.60	&	0.87	&	1.29	&	1	\\
ESO141-G055	&	1	&	0.036000	&	44.08	&	43.92	&	44.03	&	40.81	&	41.21	&	41.32	&	0	&	0	&	3.53	&	3	\\
ESO541-IG012	&	2	&	0.056552	&	44.49	&	44.42	&	44.43	&	41.13	&	41.16	&	41.56	&	0	&	0	&	5.6	&	3	\\
ESO545-G013	&	1	&	0.033730	&	44.02	&	44.26	&	44.37	&	41.40	&	41.42	&	41.46	&	54.13	&	7.91	&	3.3	&	3	\\
F01475-0740	&	2	&	0.017666	&	43.89	&	43.55	&	43.46	&	41.04	&	40.91	&	40.68	&	75.89	&	4.77	&	1.31	&	2	\\
F04385-0828	&	2	&	0.015100	&	43.98	&	43.84	&	43.78	&	41.07	&	40.95	&	40.77	&	73.51	&	5.05	&	1.12	&	2	\\
F15480-0344	&	2	&	0.030300	&	44.25	&	44.03	&	44.24	&	41.15	&	41.48	&	41.84	&	0	&	0	&	2.27	&	2	\\
I Zw 1	&	1	&	0.061142	&	45.09	&	44.98	&	44.95	&	41.74	&	42.02	&	41.55	&	75.00	&	23.92	&	18.07	&	4	\\
I Zw 92	&	2	&	0.037800	&	44.29	&	44.32	&	44.32	&	41.88	&	41.71	&	41.54	&	82.28	&	36.21	&	11.05	&	4	\\
IC4329a	&	1.2	&	0.016054	&	44.17	&	43.75	&	43.69	&	41.06	&	41.41	&	41.77	&	0	&	0	&	1.56	&	1	\\
IRAS00198-7926	&	2	&	0.072800	&	45.22	&	45.28	&	45.22	&	41.88	&	42.23	&	42.60	&	0	&	0	&	7.26	&	3	\\
IRAS00521-7054	&	2	&	0.068900	&	45.02	&	44.74	&	44.74	&	41.80	&	41.95	&	41.97	&	51.75	&	18.85	&	6.86	&	3	\\
IRASF15091-2107	&	1	&	0.044607	&	44.42	&	44.53	&	44.49	&	41.71	&	41.86	&	42.14	&	18.47	&	5.47	&	4.39	&	3	\\
IRASF22017+0319	&	2	&	0.061100	&	44.87	&	44.69	&	44.63	&	41.70	&	42.08	&	42.39	&	0	&	0	&	6.06	&	3	\\
MCG-2-40-4	&	2	&	0.025194	&	44.16	&	44.40	&	44.45	&	41.29	&	41.40	&	41.40	&	47.28	&	5.32	&	1.88	&	2	\\
MCG-2-58-22	&	1.5	&	0.046860	&	\nodata	&	\nodata	&	\nodata	&	41.28	&	41.69	&	41.70	&	5.19	&	0.58	&	4.62	&	1	\\
MCG-2-8-39	&	2	&	0.029894	&	44.05	&	43.70	&	43.83	&	41.14	&	41.29	&	41.47	&	14.66	&	1.17	&	2.24	&	2	\\
MCG-3-58-7	&	2	&	0.031462	&	44.32	&	44.42	&	44.42	&	41.04	&	41.34	&	41.42	&	2.37	&	0.15	&	2.36	&	2	\\
MCG-5-13-17	&	1.5	&	0.012642	&	43.37	&	43.38	&	43.39	&	40.58	&	40.65	&	40.62	&	38.41	&	0.85	&	0.94	&	2	\\
MCG-6-30-15	&	1.2	&	0.007749	&	43.10	&	42.85	&	42.80	&	39.71	&	39.11	&	40.41	&	0	&	0	&	0.57	&	2	\\
MRK 1034NED01	&	1	&	0.033633	&	44.32	&	44.91	&	44.95	&	41.94	&	40.96	&	40.83	&	96.03	&	48.58	&	3.29	&	3	\\
MRK 1034NED02	&	1	&	0.033710	&	\nodata	&	\nodata	&	\nodata	&	41.67	&	40.57	&	40.29	&	97.33	&	26.41	&	3.3	&	3	\\
MRK 231         &       1       &       0.04217         &       45.61   &       45.80   &       45.75   &       41.86   &       41.35   &       41.60   &       79.43
&       33.68   &       4.15    &       5       \\
MRK 273	&	2	&	0.037780	&	44.94	&	45.54	&	45.49	&	42.23	&	42.13	&	42.30	&	66.28	&	64.93	&	11.04	&	4	\\
MRK 3 	&	2	&	0.014000	&	44.17	&	43.90	&	43.85	&	41.56	&	41.94	&	41.95	&	20.13	&	4.25	&	1.04	&	2	\\
MRK 334	&	1.8	&	0.021945	&	44.12	&	44.36	&	44.31	&	41.50	&	41.44	&	41.20	&	77.74	&	14.19	&	1.64	&	2	\\
MRK 335 	&	1.2	&	0.025785	&	43.82	&	43.40	&	43.43	&	40.48	&	40.54	&	41.03	&	0	&	0	&	2.51	&	1	\\
MRK 348	&	2	&	0.015034	&	43.69	&	43.50	&	43.48	&	40.80	&	41.09	&	41.07	&	13.41	&	0.49	&	1.12	&	2	\\
MRK 471	&	1.8	&	0.034234	&	43.75	&	44.00	&	44.14	&	40.89	&	40.89	&	41.02	&	36.03	&	1.62	&	2.57	&	2	\\
MRK 509	&	1.2	&	0.034397	&	44.34	&	44.25	&	44.22	&	41.53	&	41.63	&	41.75	&	42.16	&	8.36	&	3.37	&	1	\\
MRK 573	&	2	&	0.017179	&	43.86	&	43.54	&	43.52	&	40.92	&	41.19	&	41.70	&	0	&	0	&	4.97	&	4	\\
MRK 6	&	1.5	&	0.018813	&	43.80	&	43.66	&	43.66	&	41.27	&	41.56	&	41.59	&	20.88	&	2.23	&	1.4	&	2	\\
MRK 609	&	1.8	&	0.034488	&	44.13	&	44.53	&	44.58	&	41.50	&	41.37	&	41.32	&	71.80	&	13.11	&	2.59	&	2	\\
MRK 622	&	2	&	0.023229	&	43.76	&	43.88	&	43.85	&	40.85	&	40.97	&	41.07	&	21.87	&	0.90	&	1.73	&	2	\\
MRK 79	&	1.2	&	0.022189	&	43.99	&	43.91	&	43.93	&	41.03	&	41.33	&	41.74	&	0	&	0	&	1.65	&	2	\\
MRK 817	&	1.5	&	0.031455	&	44.49	&	44.36	&	44.33	&	41.04	&	40.94	&	41.12	&	44.77	&	2.82	&	2.36	&	2	\\
MRK 883	&	1.9	&	0.037496	&	43.91	&	44.20	&	44.17	&	41.54	&	41.40	&	41.49	&	64.15	&	12.73	&	2.82	&	2	\\
MRK 897	&	2	&	0.026340	&	44.20	&	44.35	&	44.40	&	41.56	&	40.82	&	39.97	&	98.10	&	20.75	&	2.57	&	3	\\
MRK 9	&	1.5	&	0.039874	&	44.27	&	44.13	&	44.14	&	41.06	&	40.83	&	41.29	&	26.41	&	1.74	&	3.92	&	3	\\
NGC 1068	&	2	&	0.003793	&	44.50	&	44.43	&	44.42	&	41.33	&	41.69	&	41.77	&	4.34	&	0.54	&	1.09	&	4	\\
NGC 1125	&	2	&	0.010931	&	43.41	&	43.63	&	43.61	&	40.84	&	40.90	&	40.87	&	45.40	&	1.83	&	0.81	&	2	\\
NGC 1194	&	1	&	0.013596	&	43.39	&	43.19	&	43.17	&	40.30	&	40.56	&	40.75	&	0	&	0	&	1.01	&	2	\\
NGC 1275        &       2       &       0.017559        &       44.45   &       44.38   &       44.33   &       41.51   &       41.13   &       40.74   &       90.86
&       16.92   &       1.71    &       5       \\
NGC 1320	&	2	&	0.013515	&	43.70	&	43.63	&	43.62	&	40.55	&	40.55	&	41.10	&	0	&	0	&	1	&	2	\\
NGC 1365	&	1.8	&	0.005457	&	43.92	&	44.39	&	44.44	&	41.03	&	40.59	&	41.04	&	51.09	&	3.14	&	0.53	&	1	\\
NGC 1667	&	2	&	0.015167	&	43.61	&	44.17	&	44.28	&	41.11	&	40.98	&	40.78	&	75.59	&	5.67	&	1.13	&	2	\\
NGC 2639	&	1.9	&	0.011128	&	42.83	&	43.42	&	43.61	&	40.47	&	40.21	&	39.90	&	79.37	&	1.35	&	0.83	&	2	\\
NGC 2992	&	2	&	0.007710	&	43.33	&	43.68	&	43.77	&	40.84	&	40.89	&	41.12	&	12.90	&	0.52	&	0.75	&	1	\\
NGC 3079	&	2	&	0.003723	&	42.91	&	43.82	&	43.88	&	40.64	&	39.92	&	40.03	&	82.52	&	2.10	&	0.28	&	2	\\
NGC 3227	&	1.5	&	0.003859	&	42.83	&	43.10	&	43.18	&	40.32	&	40.37	&	40.29	&	39.90	&	0.48	&	0.29	&	1	\\
NGC 3516	&	1.5	&	0.008836	&	43.26	&	43.17	&	43.16	&	40.13	&	40.64	&	40.81	&	0	&	0	&	0.65	&	2	\\
NGC 3660	&	2	&	0.012285	&	42.94	&	43.48	&	43.59	&	40.33	&	39.69	&	40.07	&	60.83	&	0.75	&	1.19	&	3	\\
NGC 3783	&	1.5	&	0.009730	&	43.79	&	43.52	&	43.53	&	40.63	&	40.73	&	40.78	&	24.23	&	0.59	&	0.94	&	1	\\
NGC 3786	&	1.8	&	0.008933	&	\nodata	&	\nodata	&	\nodata	&	40.28	&	40.31	&	40.51	&	0	&	0	&	0.66	&	2	\\
NGC 3982	&	2	&	0.003699	&	42.47	&	42.98	&	43.08	&	39.67	&	39.17	&	38.77	&	84.94	&	0.23	&	0.27	&	2	\\
NGC 4051	&	1.5	&	0.002336	&	42.35	&	42.62	&	42.79	&	39.30	&	39.43	&	39.57	&	0	&	0	&	0.17	&	2	\\
NGC 4151	&	1.5	&	0.003319	&	43.14	&	42.88	&	42.88	&	40.50	&	40.77	&	40.76	&	3.42	&	0.06	&	0.25	&	2	\\
NGC 424	&	2	&	0.011764	&	43.80	&	43.43	&	43.39	&	40.65	&	40.80	&	40.73	&	35.32	&	0.92	&	0.87	&	2	\\
NGC 4388	&	2	&	0.008419	&	43.80	&	43.89	&	43.93	&	41.11	&	41.17	&	41.59	&	0	&	0	&	0.81	&	1	\\
NGC 4501	&	2	&	0.007609	&	43.28	&	44.93	&	44.15	&	39.94	&	39.77	&	39.72	&	50.98	&	0.26	&	0.74	&	3	\\
NGC 4507	&	1.9	&	0.011801	&	43.70	&	43.81	&	43.80	&	41.01	&	40.94	&	41.06	&	46.83	&	2.76	&	1.14	&	1	\\
NGC 4748	&	1	&	0.014630	&	43.31	&	43.43	&	43.48	&	40.53	&	40.87	&	41.58	&	0	&	0	&	1.42	&	3	\\
NGC 4941	&	2	&	0.003696	&	42.27	&	42.30	&	42.46	&	39.63	&	39.84	&	39.87	&	0	&	0	&	0.36	&	1	\\
NGC 4968	&	2	&	0.009863	&	43.42	&	43.40	&	43.38	&	40.70	&	40.70	&	40.80	&	34.47	&	1.00	&	0.73	&	2	\\
NGC 5005	&	2	&	0.003156	&	42.49	&	43.32	&	43.45	&	40.11	&	39.53	&	39.58	&	74.13	&	0.55	&	0.23	&	2	\\
NGC 513	&	2	&	0.019544	&	43.44	&	43.91	&	43.98	&	40.96	&	40.82	&	40.87	&	58.41	&	3.09	&	1.46	&	2	\\
NGC 5256	&	2	&	0.027863	&	44.30	&	44.80	&	44.81	&	42.11	&	41.80	&	42.02	&	74.46	&	55.83	&	2.08	&	2	\\
NGC 526A	&	1.5	&	0.019097	&	\nodata	&	\nodata	&	\nodata	&	40.63	&	40.87	&	41.18	&	0	&	0	&	1.86	&	1	\\
NGC 5347	&	2	&	0.007789	&	43.18	&	42.97	&	43.02	&	39.83	&	39.74	&	39.94	&	4.73	&	0.02	&	0.75	&	1	\\
NGC 5506	&	1.9	&	0.006181	&	43.55	&	43.54	&	43.50	&	40.62	&	40.78	&	41.21	&	0	&	0	&	0.6	&	1	\\
NGC 5548	&	1.5	&	0.017175	&	43.77	&	43.54	&	43.55	&	40.41	&	40.95	&	40.98	&	0	&	0	&	1.28	&	2	\\
NGC 5643	&	2	&	0.003999	&	43.18	&	43.52	&	43.58	&	40.19	&	40.28	&	40.60	&	0	&	0	&	1.15	&	1	\\
NGC 5929	&	2	&	0.008312	&	43.46	&	43.83	&	43.84	&	40.49	&	40.13	&	40.02	&	75.93	&	1.37	&	0.62	&	2	\\
NGC 5953	&	2	&	0.006555	&	43.11	&	43.67	&	43.72	&	40.99	&	40.29	&	40.31	&	86.57	&	4.86	&	0.48	&	2	\\
NGC 6240	&	2	&	0.024480	&	44.73	&	45.17	&	45.15	&	42.41	&	41.91	&	41.80	&	91.42	&	134.62	&	2.39	&	1	\\
NGC 6300	&	2	&	0.003699	&	42.90	&	43.33	&	43.44	&	39.65	&	39.66	&	40.06	&	0	&	0	&	0.36	&	1	\\
NGC 6890	&	2	&	0.008069	&	43.04	&	43.43	&	43.50	&	40.20	&	39.96	&	40.15	&	39.01	&	0.36	&	0.78	&	3	\\
NGC 7130	&	2	&	0.016151	&	44.16	&	44.67	&	44.69	&	41.60	&	41.32	&	41.39	&	75.04	&	17.46	&	1.2	&	2	\\
NGC 7172        &       2       &       0.008683        &       43.17   &       43.67   &       43.74   &       40.71   &       40.41   &       40.80   &       35.54
&       1.05    &       0.84    &       1       \\
NGC 7213	&	1.5	&	0.005839	&	42.81	&	42.99	&	43.14	&	40.28	&	39.99	&	39.58	&	82.88	&	0.91	&	0.56	&	1	\\
NGC 7314	&	1.9	&	0.004763	&	42.53	&	42.96	&	43.16	&	39.68	&	40.02	&	40.49	&	0	&	0	&	0.46	&	1	\\
NGC 7469	&	1.5	&	0.016317	&	44.60	&	44.87	&	44.87	&	42.10	&	41.32	&	41.58	&	88.55	&	64.57	&	1.58	&	1	\\
NGC 7582	&	2	&	0.005254	&	43.66	&	44.16	&	44.17	&	40.94	&	40.60	&	41.01	&	43.66	&	2.22	&	0.39	&	1	\\
NGC 7590	&	2	&	0.005255	&	42.77	&	43.31	&	43.49	&	39.66	&	39.32	&	39.52	&	36.31	&	0.10	&	0.51	&	3	\\
NGC 7603	&	1.5	&	0.029524	&	43.74	&	43.91	&	44.01	&	41.32	&	41.13	&	41.13	&	70.85	&	8.65	&	2.21	&	2	\\
NGC 7674	&	2	&	0.028924	&	44.62	&	44.71	&	44.72	&	41.52	&	41.93	&	41.93	&	15.63	&	2.99	&	2.16	&	2	\\
NGC 931	&	1.5	&	0.016652	&	43.98	&	43.89	&	43.93	&	40.68	&	41.12	&	41.41	&	0	&	0	&	1.24	&	2	\\
NGC 985	&	1	&	0.043143	&	44.42	&	44.46	&	44.46	&	41.47	&	41.69	&	41.74	&	33.49	&	5.72	&	4.24	&	1	\\
PG1501+106	&	1.5	&	0.036420	&	44.22	&	43.83	&	43.82	&	41.19	&	41.56	&	41.73	&	0	&	0	&	3.57	&	1	\\
PG1534+580	&	1	&	0.029577	&	43.52	&	43.22	&	43.39	&	40.64	&	40.96	&	41.07	&	0	&	0	&	2.21	&	1	\\
TOL1238-364	&	2	&	0.010924	&	43.84	&	43.99	&	43.99	&	41.00	&	40.94	&	40.49	&	81.70	&	4.75	&	0.81	&	2	\\
UGC 12138	&	1.8	&	0.024974	&	43.75	&	43.75	&	43.90	&	40.61	&	41.04	&	41.18	&	0	&	0	&	1.86	&	2	\\
UGC 7064	&	1.9	&	0.024997	&	43.79	&	44.27	&	44.34	&	40.98	&	41.14	&	41.37	&	0	&	0	&	1.87	&	2	\\
UM 146	&	1.9	&	0.017405	&	43.11	&	43.19	&	43.30	&	40.30	&	40.42	&	40.12	&	53.83	&	0.62	&	1.29	&	2	\\

\label{ir}
\enddata
\tablecomments{The sources that shows  no detectable PAH features at 6.2~$\mu$m and 11.5~$\mu$m in their spectra are: \objectname{F15480-0344},
 \objectname{IRAS00521-7054}, \objectname{MCG-2-8-39}, \objectname{MRK 3}, \objectname{MRK 9}, \objectname{NGC 3227}, 
\objectname{NGC 4941}, \objectname{NGC 526A}, \objectname{NGC 5548}, \objectname{NGC 7314} and \objectname{UM 146}. The luminosities were calculated from the fluxes using  $H_o=71{\rm kms^{-1}Mpc^{-1}}$ and  a deceleration parameter $q_o=0$ with  redshift values
 taken from NED. The last four columns show the percentage of stellar contribution to the [Ne~II] emission-line (SC); the star formation rates (SFR) derived from the stellar component of [Ne~II]; the aperture size in the dispersion direction (in kpc) and the last column shows the   references from which the emission line fluxes were obtained. Mid- and far-infrared 
continuum fluxes at 25$\micron$, 60$\micron$ and 100$\micron$ are from the {\it IRAS} \citep{1989AJ.....98..766S,1990BAAS...22Q1325M,2003AJ....126.1607S}. 
} 
\tablerefs{(1) Present calculations, (2)\cite{2007ApJ...671..124D}, (3)\cite{2007arXiv0710.4448T}, (4)\cite{2002A&A...393..821S}, (5)\cite{2005ApJ...633..706W}
} 
\end{deluxetable}

\clearpage

\begin{deluxetable}{cccccccc}
\tabletypesize{\scriptsize}
\tablewidth{0pt}
\tablecaption{Statistical Analysis Between Seyfert 1 and Seyfert 2 Galaxies}
\tablehead{ & \multicolumn{3}{c}{Seyfert 1}& \multicolumn{3}{c}{Seyfert 2}\\
 \cline{2-4} \cline{5-7}\\
 & \colhead{Measurements} && Standard& Measurements &&Standard&${\rm P_{K-S}}$\\
 &\colhead{Available} & Mean&Deviation&Available&Mean&Deviation & (\%)}
\startdata

${\rm L_{[Ne~II]}}$&39&41.0&0.1&64&40.9&0.1&79.0\\
${\rm L_{[Ne~III]}}$&39&41.0&0.1&64&40.9&0.1&22.8\\
${\rm L_{[O~IV]}}$&39&41.2&0.1&64&40.9&0.1&10.0\\
${\rm L_{25\mu m}}$&36&44.0&0.1&62&43.7&0.1&27.3\\
${\rm L_{60\mu m}}$&36&44.0&0.1&62&44.0&0.1&70.4\\
${\rm L_{100\mu m}}$&36&43.9&0.1&62&44.0&0.1&29.4\\
${\rm L_{FIR\mu m}}$&36&43.9&0.1&62&44.0&0.1&51.3\\
{\rm [O IV]/[Ne II]}&39&2.7&0.4&64&1.6&0.2&0.9\\
{\rm [Ne III]/[Ne II]}&39&1.5&0.1&64&1.03&0.08&1.1\\
{\rm [O IV]/[Ne III]}&39&2.1&0.5&64&1.4&0.1&37.7\\
{\rm [O IV]/${\rm F_{25\mu m}}$} ($10^{-3}$)&36&3.1&0.6&62&2.2&0.2&36.7\\
{\rm [O IV]/${\rm F_{60\mu m}}$} ($10^{-3}$)&36&3.9&0.7&62&1.8&0.3&0.2\\
{\rm [O IV]/FIR} ($10^{-3}$)&36&3.1&0.6&62&1.1&0.3&0.4\\
b/a                 &33&0.72&0.04&58&0.67&0.03&26.6\\
$\alpha_{25-60}\mu{\rm m}$&36&-0.8&0.1&62&-1.5&0.1&0.1\\
${\rm SC_{[Ne~II]}(\%)}$&39 &28& 5&64 & 43&4&4.4\\
${\rm SFR_{[Ne~II]}(M_\sun{\rm ~yr^{-1}})}$&39 &7& 2&64 & 8&2&18.2\\

\label{ks}
\enddata
\tablecomments{The last column, ${\rm P_{K-S}}$,  represents the Kolmogorov-Smirnov (K-S) test null probability
}  
\end{deluxetable}

\clearpage

\begin{deluxetable}{lcc}
\tabletypesize{\small}
\tablewidth{0pt} 
\tablecaption{Statistical Analysis for the Different  Relationships Between the Mid-infrared Emission Lines and Mid- and Far-infrared Continuum for the  Sample}
\tablehead{ \colhead{$\log - \log$}  &\colhead{Flux} &\colhead{Luminosity}}
\startdata 

{\rm [Ne~II]-[Ne~III]}& $r_s=0.765$&$r_s=0.853$\\\
{\rm [Ne~II]-[O~IV]} & $r_s=0.557$&$r_s=0.729$\\
{\rm [Ne~III]-[O~IV]} & $r_s=0.878$&$r_s=0.916$\\
{\rm $L_{25\micron}$-[Ne~II]}&&$r_s=0.883$\\
{\rm $L_{60\micron}$-[Ne~II]}&&$r_s=0.885$\\
{\rm $L_{100\micron}$-[Ne~II]}&&$r_s=0.796$\\
{\rm $L_{FIR}$-[Ne~II]}&&$r_s=0.856$\\
{\rm $L_{25\micron}$-[Ne~III]}&&$r_s=0.867$\\
{\rm $L_{60\micron}$-[Ne~III]}&&$r_s=0.765$\\
{\rm $L_{100\micron}$-[Ne~III]}&&$r_s=0.663$\\
{\rm $L_{FIR}$-[Ne~III]}&&$r_s=0.732$\\
{\rm $L_{25\micron}$-[O~IV]}&&$r_s=0.809$\\
{\rm $L_{60\micron}$-[O~IV]}&&$r_s=0.674$\\
{\rm $L_{100\micron}$-[O~IV]}&&$r_s=0.586$\\
{\rm $L_{FIR}$-[O~IV]}&&$r_s=0.646$\\

\enddata
\label{corre1}
\tablecomments{ $r_s$ is the
Spearman rank order correlation coefficient. The null probabilities are  $P_r < 10^{-7}$  in all these correlations.}
\end{deluxetable}

\clearpage

\begin{deluxetable}{lcccc}
\tabletypesize{\small}
\tablewidth{0pt} 
\tablecaption{Linear Regressions and Statistical Analysis for the [O~IV], [Ne~III] and [Ne~II] Fluxes and Luminosities for the 
pure AGN Sources}
\tablehead{  &\multicolumn{2}{c}{$ \log {\rm [Ne~III]}$} & \multicolumn{2}{c}{$\log {\rm [O~IV]}$}\\ 
             &\colhead{a} &\colhead{b}&\colhead{a}&\colhead{b}}
\startdata 

\multicolumn{2}{l}{{\rm Fluxes}}\\
$\log {\rm F_{{\rm [Ne~II]}}}$ & 0.81671   &-2.5689  &0.71895       &-3.9149\\
                        &$0.8\pm 0.1$&$2 \pm 1$&$0.7 \pm 0.1$&$-4 \pm 2$\\
&\multicolumn{2}{c}{$r_s=0.90$, $P_r=4.4 \times 10^{-3}$}&\multicolumn{2}{c}{$r_s=0.87$, $P_r=5.8 \times 10^{-3}$}\\
\multicolumn{2}{l}{{\rm Luminosities}}\\
$\log {\rm L_{{\rm [Ne~II]}}}$&0.86602&5.3088&0.82205&6.9797\\
                        &$0.9 \pm 0.1$&$5 \pm 4$&$0.8\pm 0.1$&$7 \pm 5$\\
&\multicolumn{2}{c}{$r_s=0.945$, $P_r=2.6 \times 10^{-3}$}&\multicolumn{2}{c}{$r_s=0.945$, $P_r=2.0 \times 10^{-3}$}\\
$\log {\rm L_{25~\micron}}$&&&1.0395&0.97779\\
                         &&&$1.0\pm0.1$&$1\pm6$\\
$\log {\rm L_{60~\micron}}$&&&0.79048&11.128\\
                         &&&$0.8\pm0.1$&$11\pm6$\\
$\log {\rm L_{100~\micron}}$&&&0.66421&16.415\\
                         &&&$0.7\pm0.1$&$16\pm6$\\
$\log {\rm L_{FIR}}$&&&0.74334&13.148\\
                   &&&$0.7\pm0.1$&$13\pm6$\\
\enddata
\label{corre}
\tablecomments{The sources that shows  no detectable PAH features at 6.2~$\mu$m and 11.5~$\mu$m in their spectra are: \objectname{F15480-0344},
 \objectname{IRAS00521-7054}, \objectname{MCG-2-8-39}, \objectname{MRK 3}, \objectname{MRK 9}, \objectname{NGC 3227}, 
\objectname{NGC 4941}, \objectname{NGC 526A}, \objectname{NGC 5548}, \objectname{NGC 7314} and \objectname{UM 146}. The regression coefficient (slope) and regression constant (intercept) are denoted by a and b, respectively. $r_s$ is the
Spearman rank order correlation coefficient and $P_r$ is the null probability. For each relation we presented the exact linear regression values,
 the values as constrained by their statistical errors and the Spearman rank and null probability.}
\end{deluxetable}

\end{document}